# Von Luchsen und Linden: Neue Einblicke in die historische Ökologie des Bayerischen Waldes an der Schwelle zur Moderne

Bettina Haas, Markus Gerstmeier, Ricarda Huter, Malte Rehbein

## Überblick

Am Ende der „Sattelzeit" stand auch das Königreich Bayern mit seinen z. T. bemerkenswert frühen Ansätzen zur Rationalisierung der Staatsverwaltung, auch zur Industrialisierung und mit seinen Erschließungsprojekten nicht nur auf dem Gebiet der Infrastruktur an der Schwelle zur Moderne. In diesen Kontext[1] gehört die in der Forschung bislang wenig beachtete wissenschaftliche Erfassung der Flora und Fauna im gesamten Königreich Bayern im Jahr 1845.[2] Die vorliegende Studie möchte anhand eines exemplarischen bayerischen Forstamtes, nämlich des für den inneren Bayerischen Wald zuständigen Forstamtes Zwiesel und anhand des in der Erhebung von 1845 dokumentierten Tier- und Baumbestandes in dessen Gebiet verschiedene Perspektiven auf die historische Ökologie eröffnen: auf Forstwesen und Forstverwaltung,[3] auf das Vorkommen von Tier- und Pflanzenarten in der Mitte des 19. Jahrhunderts, auf die menschliche Rezeption der natürlichen Umwelt und auf menschliches Einwirken auf die Natur (und umgekehrt), außerdem, in nicht zuletzt wissenschaftshistorischer Hinsicht, auf zeitgenössische Ansätze systematischer Erschließung von Biodiversität.

---

[1] Bernhard Löffler nennt in diesem Zusammenhang das zweite Kapitel des ersten Teils seines viel beachteten Werkes über „Das Land der Bayern. Geschichte und Geschichten von 1800 bis heute", München 2024, S. 31-86, nicht von ungefähr „Vermessung und Vermessenheit: Die kartographisch statistische Erfindung des Raumes Bayern".

[2] Zum Überblick vgl. Malte Rehbein, From Historical Archives to Algorithms: Reconstructing Biodiversity Patterns in 19th Century Bavaria, in: Diversity 17, Nr. 5 (26. April 2025): 315. https://doi.org/10.3390/d17050315.

[3] Anknüpfend an Thorsten Franz, Geschichte der deutschen Forstverwaltung, Wiesbaden 2020; Richard Hölzl, Umkämpfte Wälder: Die Geschichte einer ökologischen Reform in Deutschland 1760–1860, Frankfurt am Main 2010.

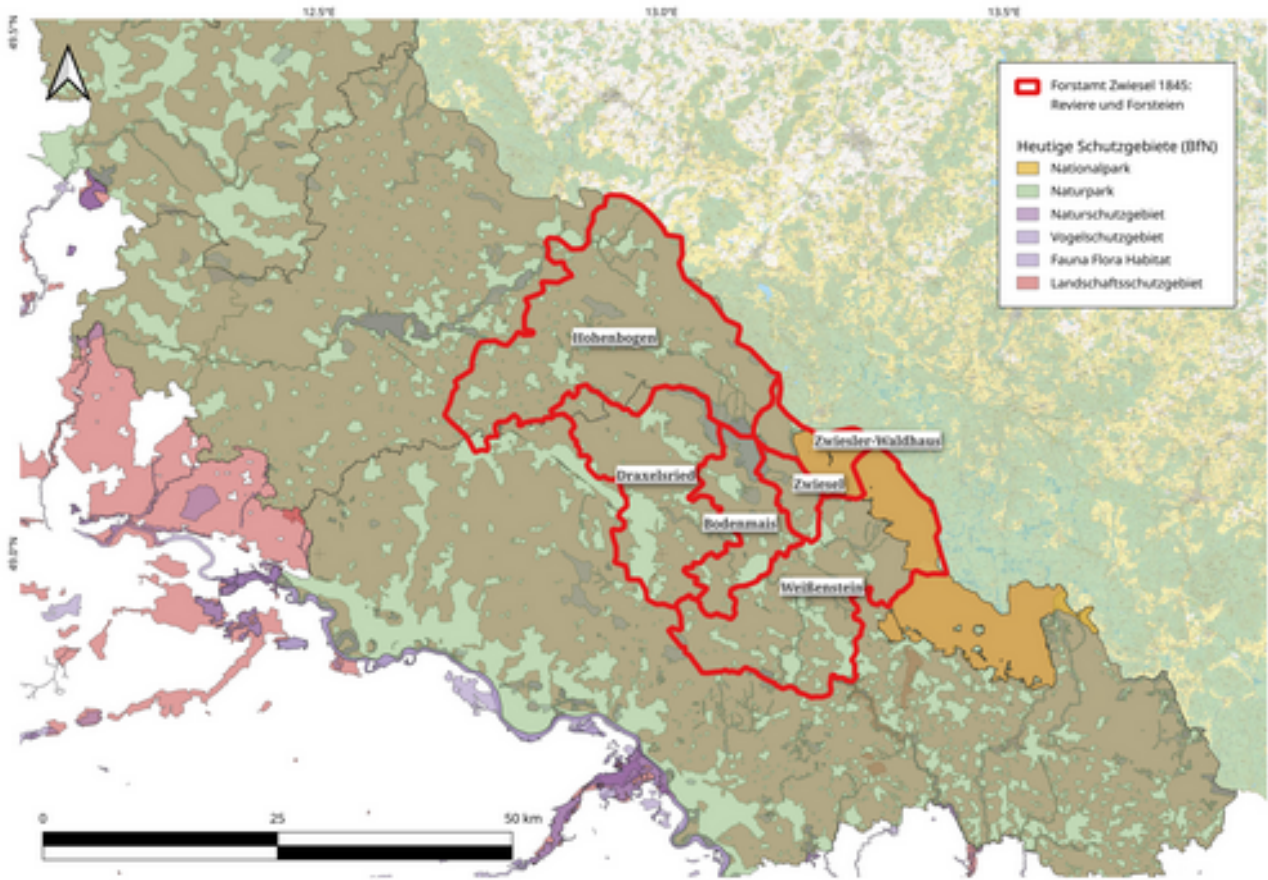

*Abbildung 1. Zuständigkeitsbereich des Forstamts Zwiesel mit dessen untergeordneten Revieren Bodenmais, Draxelsried, Hohenbogen Zwiesel und Zwiesler-Waldhaus sowie der Forstei Weißenstein um 1845. Die Rekonstruktion der Reviergrenzen erfolgte mit Hilfe der „Forstlichen Uebersichtskarte des Königreichs Bayern diesseits des Rheins in sechs Blättern", die 1857 im Königlich Ministerialen Forsteinrichtungs Bureau, vermutlich in Zusammenhang mit einer Verwaltungsreform, entstand (Staatsarchiv Amberg, Forstdirektion Niederbayern-Oberpfalz Kartensammlung 3/1-3). Wir danken Silke Dutzmann vom Leibniz-Institut für Länderkunde (IfL) Leipzig für die Erstellung der shape-Files.*

## Zoologie und Botanik, Flora und Fauna in Bayern 1845

Die Mitte des 19. Jahrhunderts markierte für das Königreich Bayern eine Phase des Wandels. Im Kontext der Revolution von 1848/49 kam es hier zum Rücktritt des schönheits- und kunstbeflissenen Königs Ludwig I. (reg. 1825–1848) zugunsten seines Sohnes Maximilian II. Joseph (reg. 1848–1864).[4] Parallel zur politischen Transformation erlebten die von König Maximilian II. in erheblichem Maße geförderten Geistes- und Naturwissenschaften eine Phase bedeutender Fortschritte. In diesem Umfeld wurde in Zoologie und Botanik an die Blüte der systematischen Sammlungs-, Beschreibungs und Klassifizierungstätigkeiten angeknüpft, für die während des 18. Jahrhunderts im Zeichen der Aufklärung so bekannte Forscherpersönlichkeiten wie Carl von Linné (1707–1778) gestanden hatten. Auch durch die Impulse von Entdeckungsreisenden wie Alexander von Humboldt (1769–1859) bildete sich im Laufe des 19. Jahrhunderts immer weiter ein umfassendes naturwissenschaftliches Weltbild heraus; andersherum erzählt, kam es auf diesem Weg jetzt überhaupt erst, von menschlicher Wahrnehmungsverschiebung her, zur „Erfindung der Natur", wie Andrea Wulf ihre Biographie Alexander von Humboldts treffend überschreibt.[5] Die darauf aufbauenden Arbeiten Charles Darwins (1809–1882), etwa seine "Zoology of the Voyage of H.M.S. Beagle Under the Command of Captain Fitzroy, R[oyal] N[avy], during the Years 1832 to 1836" (5 Bde., London 1838–1843) mündeten schließlich 1859 – freilich erst nach den in diesem Beitrag beschriebenen Ereignissen und Quellen – in die von ihm und parallel von Alfred Russel Wallace (1823–1913) entwickelte Evolutionstheorie.

---

[4] Hierzu Marita Krauss, König Ludwig I. von Bayern − Träume und Macht. Eine Biografie, München 2025, S. 427-459.

Im bayerischen Kontext ragt hier Carl Philipp von Martius (1794–1868) hervor, der auf einer im Auftrag des ersten Königs von Bayern, Maximilians I. Joseph (reg. 1799/1806–1825), unternommenen Forschungsreise nach Brasilien zusammen mit Johann Baptist von Spix (1781–1826) die Grundlage für die von ihm begründete Reihe "Flora Brasiliensis" schuf, die von 1840 an erschien. Martius wiederum führt uns zu Johann Andreas Wagner (1797–1861), dessen verglichen mit Darwin – bei ähnlichen Ausgangspunkten, nämlich dem in beider akademischen Sozialisationen dies- und jenseits des Ärmelkanals noch weithin vorherrschenden Primat der scholastischen Theologie, also thomistischer Deduktion in aristotelischer Tradition anstelle von moderner induktiver Empirie auf dem akademischen Feld – persistent religiös-christliche Hauptmotivation[6] für Naturforschung gewiss auch die

---

[5] Andrea Wulf, The Invention of Nature: How Alexander Von Humboldt Revolutionized Our World, New York, N.Y., 2015; deutsch: Alexander von Humboldt und die Erfindung der Natur, München 2016.

[6] Hierzu eingehend Andrea Alaoui Soulimani, Naturkunde unter dem Einfluss christlicher Religion. Johann Andreas

Komplexität wissenschaftlichen Fortschritts verdeutlicht. Beide stammten nicht nur aus den unter Napoleonischen Vorzeichen an Bayern gefallenen Teilen Frankens – von Martius aus dem vormals brandenburgisch-ansbachisch-bayreuthischen Erlangen, Wagner aus der altehrwürdigen freien Reichsstadt Nürnberg –, sondern waren auch Kollegen in der Philosophischen Fakultät der Ludwig-Maximilians-Universität,[7] an welcher Martius bei deren Verlegung nach München 1826 zum ordentlichen Professor für Mineralogie avanciert war. Wagner amtierte ebendort seit 1833/36 erst als außerordentlicher, dann als ordentlicher Professor für Zoologie und Naturgeschichte, fungierte außerdem bereits seit 1832 als Zweiter Konservator der Zoologisch-Zootomischen Staatssammlung in München und seit 1843 als Erster Konservator der aus dieser ausgegliederten Paläontologischen Staatssammlung.[8] 1862 wird von Martius als Sekretär der mathematisch-physikalischen Klasse der Königlich Bayerischen Akademie der Wissenschaften eine Gedenkrede zu Ehren Wagners halten.[9]

Das ist der Hintergrund von Männern wie Andreas Wagner und Carl Philipp von Martius: eine sich herausbildende naturwissenschaftliche Systematik und die strukturierte „Vermessung der Welt", so der Titel von Daniel Kehlmanns gelehrtem Roman über Alexander von Humboldt von 2005; zugleich auf globaler und regionaler Ebene sowie im Angesicht der zeitgenössischen politischen Umwälzungen der sukzessive, aber zentralistisch organisierte Aufbau einer modernen Staatsverwaltung unter anderem auch durch statistische und vermessende Methoden ebenso wie identitätsstiftende Maßnahmen

---

Wagner (1797–1861). Ein Leben für die Naturkunde in einer Zeit der Wandlungen in Methode, Theorie und Weltanschauung, Düren 2001.

[7] Verzeichnis der an der königlichen Ludwig-Maximilians-Universität zu München im Winter-Semester 1836/37 zu haltenden Vorlesungen, München o. J., S. 13 und 15.

[8] Zu den Daten und Kontexten des Werdegangs und Wirkens Wagners siehe den neueren Artikel von Birgit Hoppe, Wagner, Andreas, in: Neue Deutsche Biographie 27 (2020), S. 226 f., also im bedeutendsten biographischen Lexikon des deutschen Sprachraumes.

[9] Carl Friedrich Philipp von Martius, Denkrede auf Joh. Andreas Wagner. Gehalten in der öffentlichen Sitzung [der Königlich Bayerischen Akademie der Wissenschaften] am 28. November 1862, München 1862.

durch regionale und nationale natur- und volkskundliche Darstellungen.[10] Auch und gerade aus diesem Geist heraus ließ das Königlich Bayerische Ministerium der Finanzen im Jahr 1845 förmlich „Auf Seiner Majestät des Königs Allerhöchsten Befehl", de facto wohl aber unter maßgeblicher Förderung des wissenschaftsaffinen Kronprinzen Maximilian Joseph, eine umfassende Erhebung zur wissenschaftlichen Dokumentation der geografischen Verbreitung von Tier- und Baumarten durchführen. Die wissenschaftliche Leitung des faunistischen Teils dieser Erhebung übernahm eben Andreas Wagner. Die akademisch herausgehobene Position Wagners als Ordinarius der ersten Universität des Königreichs exemplifiziert nicht nur seine persönliche Expertise und Bedeutung in der Fachwelt, sondern weist auch darauf hin, dass der Erhebung vonseiten des Königreichs ein hoher Stellen zugerechnet worden ist. Erste Ergebnisse seiner Untersuchungen präsentierte Wagner im Frühjahr 1846 im Rahmen einer anschließend auch publizierten Vortragsreihe vor der Bayerischen Akademie der Wissenschaften , welcher er seit 1835 als außerordentliches und seit 1842 als ordentliches Mitglied angehörte.[11] Dort beschrieb er den Grund der Untersuchungen: Er sei „von Seiner Königlichen Hoheit dem Kronprinzen von Bayern mit dem Auftrage beehrt [worden,] auf einer größeren Karte eine Darstellung der geographischen Verbreitung der wichtigsten Thiere aus der bayerischen Fauna zu versuchen".[12] Die daraus entstandene „Übersichtskarte der Verbreitungsverhältnisse der merkwürdigsten wildlebenden Thiere in Bayern" blieb bislang nach unserer Erkenntnis unbeachtet.[13]

---

[10] Teile der Wagner'schen Erhebung bzw. seiner Zusammenfassung flossen in die Naturwissenschaftlichen Darstellungen Bayerns in der „Bavaria", der landes- und volkskundlichen Beschreibung des Königreich Bayern, ein.

[11] Wagner wurde im Jahr 1857 außerdem die Akademie der Naturforscher Leopoldina in Halle (Saale) gewählt.

[12] Andreas Wagner, Beyträge zur Kenntniß der bayerischen Fauna, in: Gelehrte Anzeigen, herausgegeben von Mitgliedern der königlich bayerischen Akademie der Wissenschaften 22 (1846), hier S. 698.

[13] Uebersichtskarte der Verbreitungsverhältnisse der merkwürdigsten wildlebenden Thiere in Bayern. Ein Versuch, nach Angabe königlicher Forstämter und eigner Erfahrung bearbeitet von Wagner, Doctor und Professor. München, Bayerische Staatsbibliothek, Cod.icon. 180 rc.

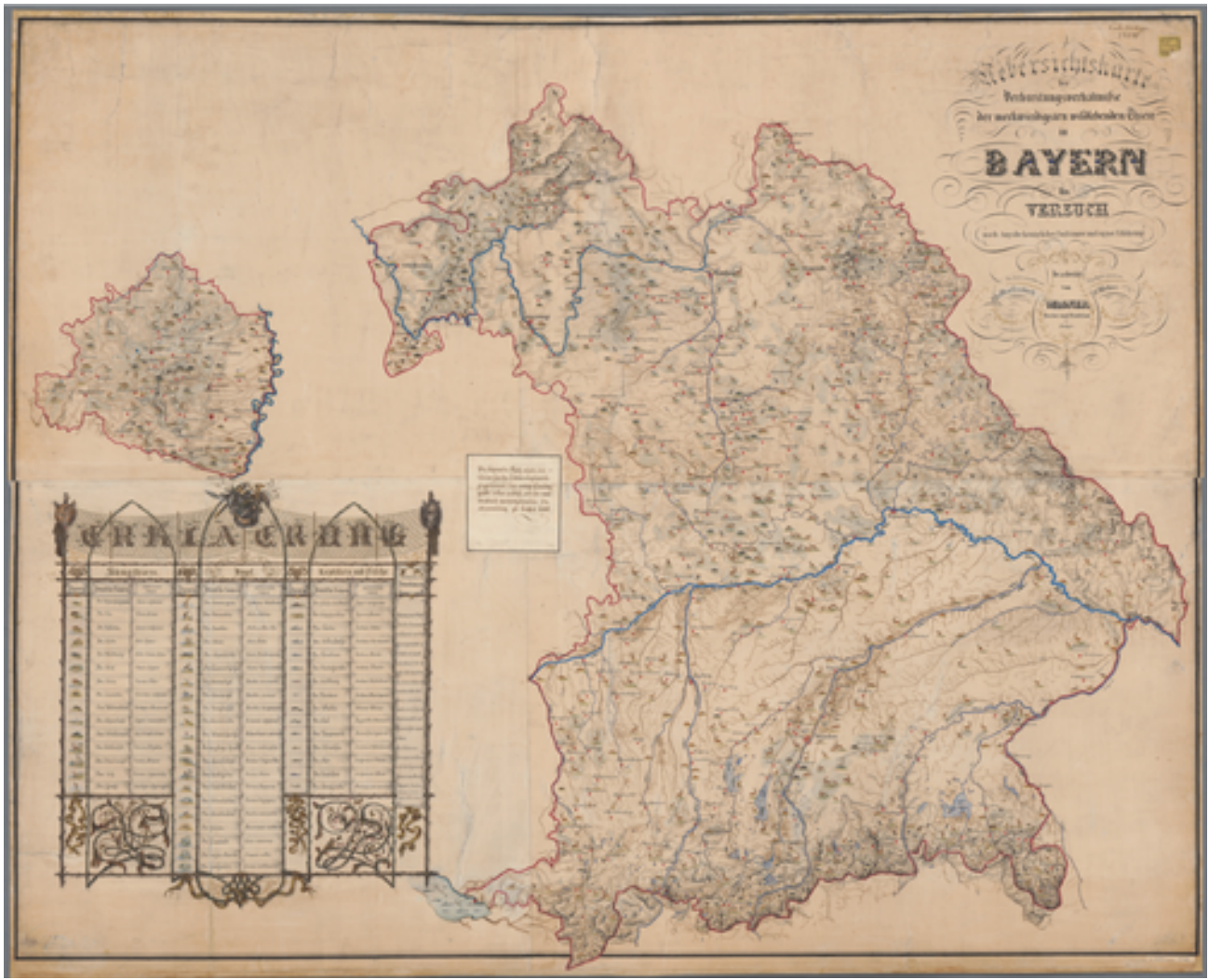

*Abbildung 2. Uebersichtskarte der Verbreitungsverhältnisse der merkwürdigsten wildlebenden Thiere in Bayern. Ein Versuch, nach Angabe königlicher Forstämter und eigner Erfahrung bearbeitet von Wagner, Doctor und Professor. – München, Bayerische Staatsbibliothek, Cod.icon. 180 rc.*

## Das Quellenmaterial

Diese von Andreas Wagner „nach Angabe königlicher Forstämter und eigner Erfahrung" angefertigte Karte ist nur eine naturhistorische Informationseinheit innerhalb des heterogenen Quellenmaterials, welches aus den Kontexten des königlich bayerischen Erhebungsprojektes zu Tier- und Pflanzenarten aus des Jahr 1845 erhalten ist. Neben der kartographischen Quelle stehen uns vor allem die an alle bayerischen Forstämter verschickten und von diesen ausgefüllten Fragebögen zur Verfügung, außerdem der Gesamtbericht, den Wagner neben und mit seiner „Uebersichtskarte" für die Regierung in München erstellt hat; des Weiteren Wagners bereits erwähnte, auf der Erhebung von 1845 basierende Beiträge für das wissenschaftliche Fachpublikum.[14] Als zusätzliche „Quellen hinter den (eigentlichen) Quellen" kommen bei Andreas Wagner außerdem Naturbeobachtungen Dritter hinzu, die er in seine Gesamtauswertung der Erhebung mit einfließen ließ – wie etwa im Falle des von uns im Folgenden zu untersuchenden Forstamtes Zwiesel Zeugnisse des gut zwei Alterskohorten vor Wagner wirkenden, in der Region verwurzelten Aristokraten Iganz Dominikus von Poschinger (1747–1803). Aufseiten der erhebenden Forstämter vor Ort werden uns als „Zusatzquellen" die bezüglich örtlicher Flora und Fauna inhaltlich umfassenderen und tiefenschärferen Auskünfte der nachgeordneten Dienststellen des Forstamtes Zwiesel, also von dessen Revieren und Forstei, begegnen, Quellen, auf deren Grundlage das Forstamt dann seine Angaben in die Haupt- und Residenzstadt München schicken sollte.

Wagners Vorgehen mithilfe eines Fragebogens war hierbei nicht ungewöhnlich für eine Periode zunehmender Empirie und Systematisierung in Wissenschaft und Staatsverwaltung. Die bayerischen Forstämter erhielten bei der Erhebung von 1845 standardisierte Fragebögen in zwei Varianten: Im Schema A – das „Verzeichnis der Thierarten, von deren Vorhandenseyn u. Wohnort Nachricht gewünscht wird"[15] – sollten die Forstämter das Vorkommen von 44 verschiedenen Wirbeltierarten erfassen, während das Schema B unter dem Titel „Fragen über die Verbreitung der in Bayern einheimischen Baumarten"[16] sieben Aspekte zum Vorkommen von Baumarten abfragte.

Die ausgefüllten Fragebögen zum Schema A verblieben, zusammen mit einigen Notizen

---

[14] Andreas Wagner, Beyträge zur Kenntniß der bayerischen Fauna, in: Gelehrte Anzeigen, herausgegeben von Mitgliedern der königlich bayerischen Akademie der Wissenschaften 22 (1846), S. 649-656, 657-664, 665-672, 673-680, 697-700.

[15] Staatsarchiv Landshut (im Folgenden: StALa), Forstamt Zwiesel A 988, S. 22.

[16] StALa, Forstamt Zwiesel A 988, S. 25.

Wagners, zunächst in der Zoologischen Staatssammlung in München, bevor sie 2013 an das Bayerische Hauptstaatsarchiv abgegeben wurden. Im Jahr 2024 wurden sie im Rahmen eines kooperativen Forschungsprojekts der Generaldirektion der Staatlichen Archive Bayerns, des Deutschen Zentrum für intergrative Biodiversitätsforschung (iDiv) Halle-Jena-Leipzig und des Lehrstuhls für Computational Humanities der Universität Passau katalogisiert, digitalisiert, verdatet und computergestützt ausgewertet, wobei über fünftausend distinkte Daten erfasst und als OpenData zugänglich gemacht wurden.[17] Es handelt sich hierbei um ein erstes Pilotprojekt eines neuen Forschungsschwerpunktes „Computational Historical Ecology" (CHE).[18]

---

[17] Rehbein, M., Escobari Vargas, A. B., Fischer, S., Güntsch, A., Haas, B., Matheisen, G., Perschl, T., Wieshuber, A., & Engel, T. (2024). Historical Animal Observation Records by Bavarian Forestry Offices (1845) (1.3) [Data set]. Zenodo. https://doi.org/10.5281/zenodo.14008158.

[18] Malte Rehbein (September 30, 2024). On Computational Historical Ecology, Part I. Computational Historical Ecology. Retrieved May 17, 2025 from https://che.hypotheses.org/69.

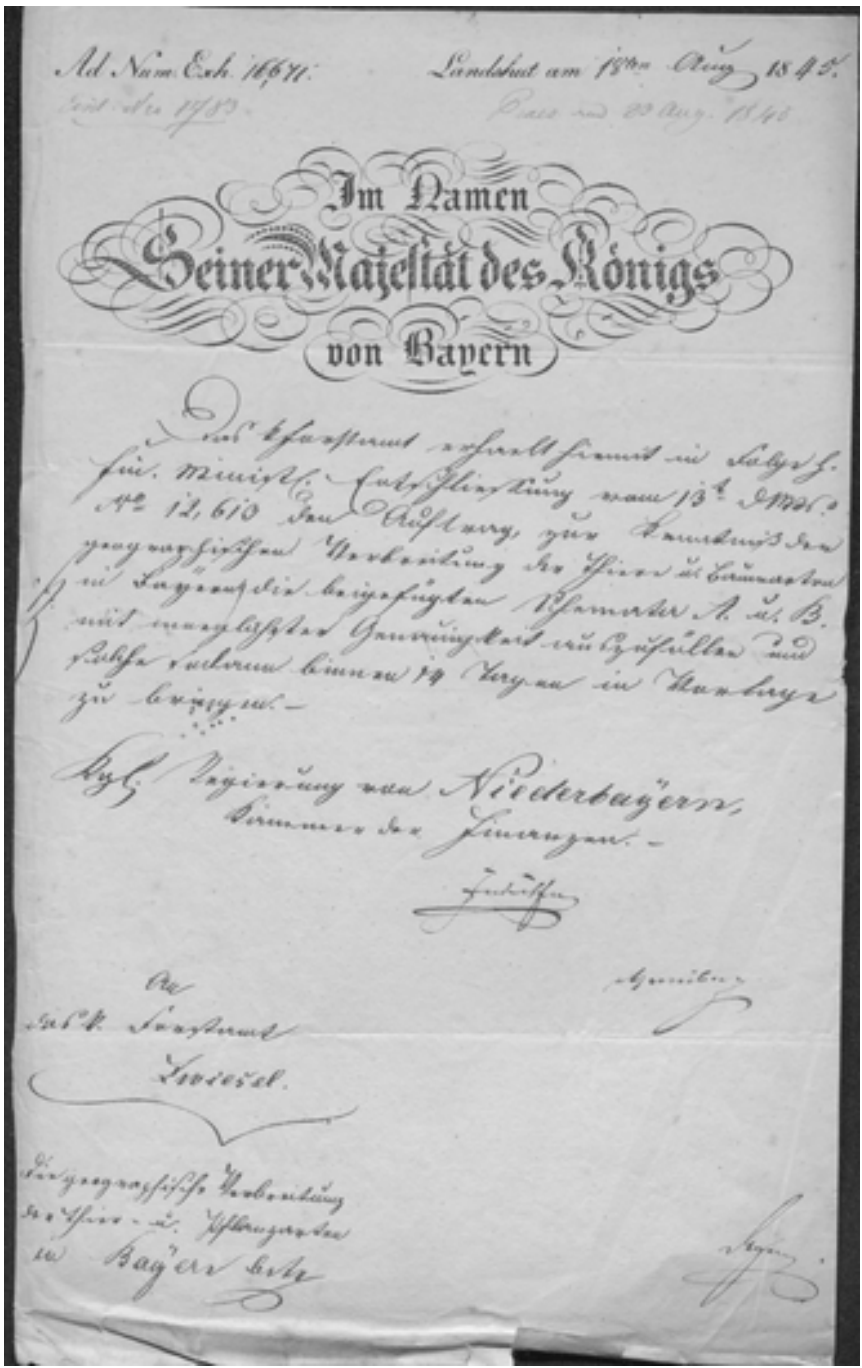

*Abbildung 3. Schreiben der königlichen Regierung von Niederbayern an das königliche Forstamt Zwiesel vom 18. August 1845 (StALa Forstamt Zwiesel A 988, S. 2).*

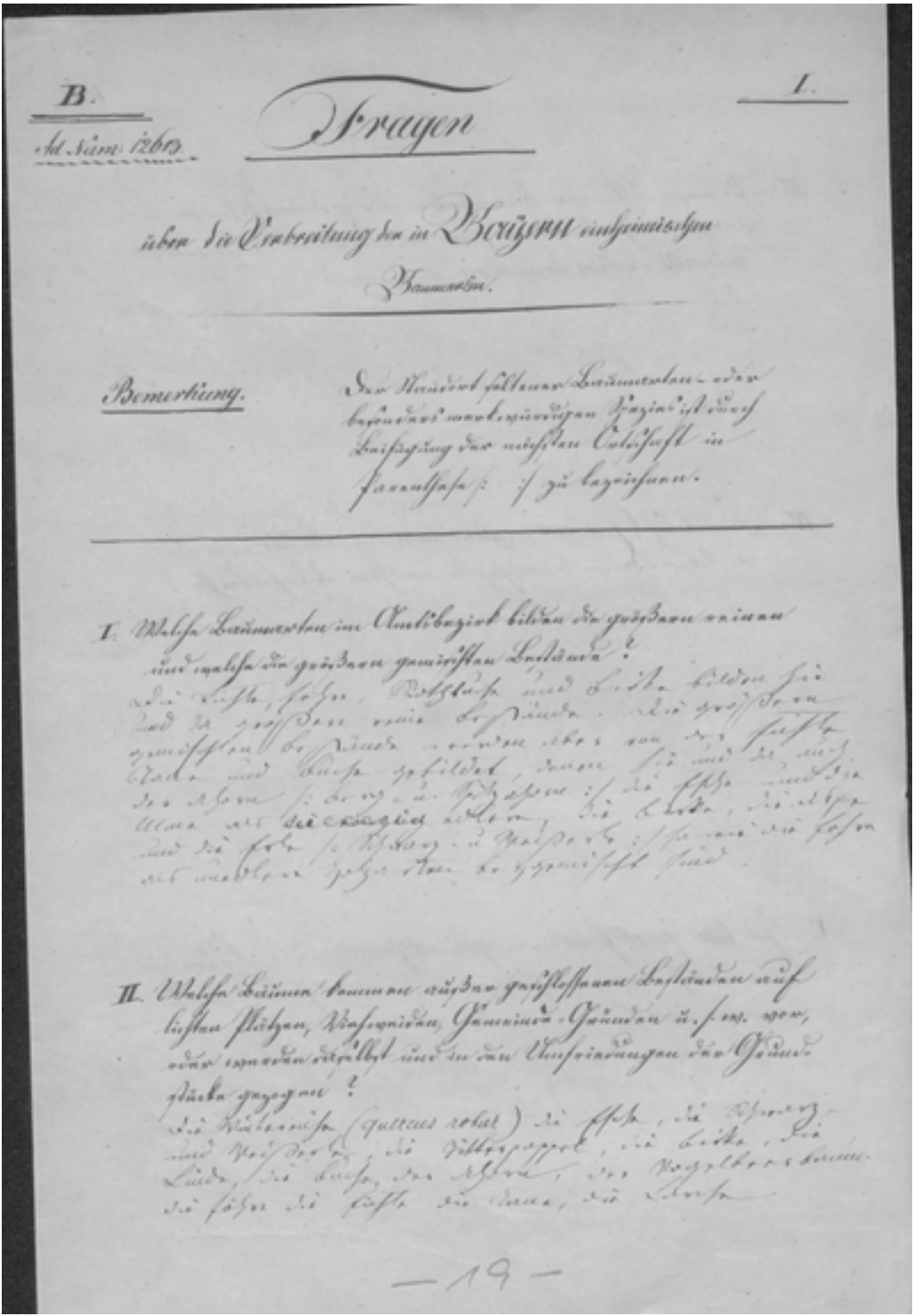

*Abbildung 4. Seite 1 des ausgefüllten Schema B des Forstamtes Zwiesel (StALa Forstamt Zwiesel A 988, S. 25).*

Demgegenüber kann über den archivalischen Verbleib von Schema B, den „Fragen über die Verbreitung der in Bayern einheimischen Baumarten"[19] weithin nur gemutmaßt werden – ein vollständiger Bestand wie bei Schema A konnte bislang nicht aufgefunden werden. Ebenso konnte bislang nicht geklärt werden, ob es eine ähnliche wissenschaftliche Betreuung und Organisation – etwa durch die Botanische Staatssammlung, die 1845 unter der Leitung von Martius' stand – dieses botanischen Anteils der Erhebung gegeben hat. Es liegen aber Hinweise darauf vor, dass die botanische Erhebung auch tatsächlich durchgeführt worden ist. Diese Spuren führen uns in den Bayerischen Wald.

## „Naturhistorische Forschungen und Beobachtungen" im Forstamt Zwiesel

Die Überlieferung der niederbayerischen Forstämter aus der Zeit des Königreichs Bayern befindet sich im Staatsarchiv Landshut. Der dortige Bestand des Forstamtes Zwiesel enthält einen umfangreichen Akt „Naturhistorische Forschungen und Beobachtungen"[20]. Dieses Konvolut umfasst verschiedene systematische Untersuchungen aus den Jahren 1845 bis 1869 etwa zum Vorkommen von Eisenerz und Bodenarten, von Holzgewächsen, zur Temperatur von Quellen, zum Vorkommen der Haselfichte sowie von sämtlichen jagdbaren Vögeln und zu klimatischen Verhältnissen.

Die Seiten 2–28 des besagten Akts sind deshalb von besonderem Interesse, weil sie die Zwieseler Daten zur Erhebung des Tier- und Baumvorkommens im Jahr 1845 dokumentieren. In verwaltungsgeschichtlicher Hinsicht ermöglicht diese Quelle einen genaueren Einblick in die Aggregation von Informationen der Forstämter durch die Beteiligung der untergeordneten Forstreviere und Forstei. Sie lässt damit zunächst Rückschlüsse auf die Arbeitsweise eines Forstamtes beim Beobachten und Erfassen der Tier- und Baumbestände in dessen Zuständigkeitsbereich zu. Weil forstliche Datenerhebungen bislang nur auf der Aggregationsebene Forstamt und damit für einem deutlich größeren geographischen Raum bekannt waren, bietet der Zwieseler Datensatz im

---

[19] StALa, Forstamt Zwiesel A 988, S. 25.

[20] StALA, Forstamt Zwiesel A 988: „Naturhistorische Forschungen und Beobachtungen".

Staatsarchiv Landshut außerdem eine detailliertere Tiefenschärfe historischer ökologischer Vigilanz.[21] Weiterhin liefert der Akt „Naturhistorische Forschungen und Beobachtungen" bisher unveröffentlichte Daten zum Schema B der Erhebung von 1845, die im Anhang des vorliegenden Beitrags veröffentlicht werden.

Das Forstamt Zwiesel bildete im Jahr 1845 eines von insgesamt acht Forstämtern im damaligen Kreis Niederbayern – neben der Forstverwaltung Deggendorf, den Forstämtern Freising[22], Kelheim, Neustadt a.D. in Geisenfeld, Passau, Schönberg und Wolfstein. Hinzu kam in der Region noch das außerhalb der staatlichen Königlich Bayerischen Forstverwaltung organisierte, bis 1849 standesherrliche Forstamt des Fürsten von Thurn und Taxis mit Sitz in Wörth an der Donau.

| **Organisationseinheit** | **Leitung** |
| --- | --- |
| Forstamt Zwiesel | Forstmeister Hr. Ferd. Klein |
|     Revier zu Bodenmais | Revierförster Hr. Karl Frhr. v. Asch |
|     Revier zu Draxelsried | Revierförster Hr. Georg Herrmann |
|     Revier zu Hohenbogen | Revierförster Hr. Friedrich v. Krafft-Festenberg |
|     Revier zu Zwiesel | Revierförster Hr. Max Ney |
|     Revier zu Zwiesler=Waldhaus | Revierförster Hr. Frz. Xaver Seninger |
|     Forstei zu Weißenstein | Forsteiförster Hr. Heinrich Ant. Pfisterer |

*Tabelle 1: Organisatorische Struktur des Forstamts Zwiesel und dessen Leitungspersonal im Jahr 1845 gemäß Hof= und Staats=Handbuch des Königreichs Bayern 1845, München o. J., S. 261.*

---

[21] Die sowohl als Forschungsdatensatz auf Zenodo als auch über die Plattform GBIF publizierten Daten beziehen sich auf die 119 bayerischen Forstämter jeweils als Ganzes. Damit ist etwa die Frage, ob eine bestimmte Tierart vorkommt, im Falle des Forstamts Zwiesel einem geographischen Raum von ca. 1276 km$^2$ zugeordnet (die Berechnung erfolgte auf Grundlage der rekonstruierten Reviergrenzen mit Hilfe der Geographischen Informationssystems QGIS: Hohenbogen: 460, Weißenstein: 346, Draxelsried: 250, Bodenmais: 119, Zwiesler-Waldhaus: 64, Zwiesel: 37 km$^2$).

[22] Zwei dem oberbayerischen Forstamt Freising zugeordnete Reviere (das Revier Siebensee und die Forstei Eberspoint) befanden sich in Niederbayern. Für den niederbayerischen Teil wurde ein eigener Fragebogen ausgefüllt.

Die in der Akte enthaltenen Unterlagen zeigen, dass das Forstamt Zwiesel die Anordnung zur Bearbeitung der Erhebungen an die ihm untergeordneten Organisationseinheiten, fünf Forstreviere und eine Forstei, weiterleitete. Deshalb finden sich in der Akte auch die ausgefüllten Fragebögen der einzelnen Forstreviere. Lediglich aus dem Forstrevier Draxelsried sind keine Antworten überliefert. Auf dieser differenzierten Datengrundlage kann eine detaillierte Bestandsaufnahme der historischen Tier- und Baumvorkommen in der Bayerwaldregion des mittleren 19. Jahrhundert vorgenommen werden. Die Verbindung von lokaler Erhebung und übergeordneter wissenschaftlicher Auswertung macht den Bestand zu einer aufschlussreichen Quelle für die historische Ökologie und landesgeschichtliche Forstkunde. Gewiss wurden und werden im 20. und 21. Jahrhundert deutlich umfassendere Bestandsaufnahmen vorgenommen.[23] Die Erhebung von 1845 und die weiteren noch aufzuarbeitenden historischen Quellen dienen gleichwohl unter anderem dazu, Entwicklungsdynamiken weiter in die Vergangenheit zurückzuverfolgen als es bisher möglich war.

## Tiervorkommen im Forstamt Zwiesel: Schema A

Die Erhebungen durch die Forstmeister der einzelnen Reviere des Forstamtes Zwiesel bieten eine wertvolle Grundlage, um das Tiervorkommen im Bayerischen Wald sowie die Perspektiven der Forstämter genauer zu beleuchten. Im Fokus dieses Abschnitts stehen die Berichte zum Tiervorkommen des Forstamts Zwiesel und seinen Revieren, die detailliert dokumentieren, welche Tierarten 1845 in den einzelnen Revieren vorkamen oder fehlten (Tab. 2). Diese Dokumente der einzelnen Reviere erlauben uns genauere Analysen über die Fauna der Region anzustellen.

---

[23] Siehe z. B. Marco Heurich / Markus Neufanger, Die Wälder des Nationalparks Bayerischer Wald: Ergebnisse der Waldinventur 2002/2003 im geschichtlichen und waldökologischen Kontext. Nationalpark Bayerischer Wald, Heft 16. Grafenau: Nationalparkverwaltung Bayerischer Wald, 2005.

| Tierart | | Revier Zwiesler Waldhaus | Revier Hohenbogen | Revier Bodenmais | Revier Zwiesel | Forstei Weißenstein |
|---|---|---|---|---|---|---|
| **Säugetiere** | ***Mammalia*** | | | | | |
| Dachs | *Meles meles* | selten | -- | sehr selten | selten | -- |
| Steinmarder | *Martes foina* | selten | alltenhalben | mittelmäßig | -- | alltenhalben |
| Edelmarder | *Martes martes* | ziemlich häufig | häufig | mittelmäßig | sehr häufig | alltenhalben |
| Fischotter | *Lutra lutra* | selten | nicht selten | einige | keine Seltenheit | alltenhalben |
| Luchs | *Lynx lynx* | -- | -- | wenige | -- | -- |
| Edelhirsch | *Cervus elaphus* | -- | nicht selten, durchwandernd | -- | -- | selten als Wechselwild |
| Reh | *Capreolus capreolus* | häufig | häufig | -- | häufig | nicht häufig, wechselweise[24] |
| **Vögel** | ***Aves*** | | | | | |
| Uhu | *Bubo bubo* | -- | -- | wenige | -- | -- |
| Saatkrähe | *Corvus frugilegus* | -- | nicht sehr häufig | -- | häufig | -- |
| Alpendohle | *Pyrrhocorax graculus* | -- | -- | mittelmäßig | -- | -- |
| Nachtigall | *Luscinia megarhynchos* | -- | selten | -- | -- | -- |
| Auerhuhn | *Tetrao urogallus* | nicht gar häufig | nicht selten | wenig | häufig | geringe Anzahl |
| Birkhuhn | *Tetrao tetrix* | -- | nicht häufig | sehr selten | sehr selten; vor einem Jahrzehnt häufig | gering |
| Haselhuhn | *Tetrastes bonasia* | ziemlich häufig | sehr häufig | mittelmäßig | [häufig] | häufiger |
| Schnepfe | *Scolopax rusticola* | ziemlich häufig | sehr häufig | mittelmäßig | sehr häufig | allenthalben, nicht häufig brütend |
| Bekassine | *Gallinago gallinago* | -- | -- | strichweise | -- | strichweise in geringer Anzahl |
| Gemeine Wildgans | *Anser anser* | -- | -- | -- | sehr selten | -- |
| Enten | *anas pp.* | -- | -- | -- | sehr selten | nicht häufig |
| **Reptilien** | ***Reptilia*** | | | | | |
| Kupferotter | *Vipera berus* | -- | -- | sehr wenige | -- | -- |

*Tabelle 2. Gemeldete Vorkommen der Tierarten mit Häufigkeitsangaben im Forstamt Zwiesel.*

Von den 44 Wirbeltierarten, die im Schema A der forstlichen Erhebung von 1845 abgefragt

---

[24] Unklar, worauf er sich mit „sämtliche vorstehende Thierarten" beziehen könnte.

Haas/Gerstmeier/Huter/Rehbein: Von Luchsen und Linden (preprint)

wurden, wurden folgende im Gebiet des Forstamtes Zwiesel als *nicht* vorkommend erfasst: Bär (*Ursus arctos*), Wildkatze (*Felis silvestris silvestris*), Wolf (*Canis lupus*), Biber (*Castor fiber*), Hamster (*Cricetus cricetus*), Murmeltier (*Marmota marmota*), Wildschwein (*Sus scrofa*), Damhirsch (*Dama dama*), Gemse (*Rupicapra rupicapra*), Lämmergeier (Bartgeier, *Gypaetus barbatus*), Steinadler (*Aquila chrysaetos*), Seeadler (*Haliaeetus albicilla*), Fischaar (Fischadler, *Pandion haliaetus*), Steinkrähe (*Pyrrhocorax graculus*), Mauerspecht (Mauerläufer, *Tichodroma muraria*), Kranich (*Grus grus*), Rohrdommel (*Botaurus stellaris*), Weißer Storch (*Ciconia ciconia*), Schwarzer Storch (*Ciconia nigra*), Höckerschwan (*Cygnus olor*), Singschwan (*Cygnus cygnus*) und Saatgans (*Anser fabalis*). Heute sind im Gebiet des Nationalparks Bayerischer Wald einige der 1845 nicht nachgewiesenen Arten wieder anzutreffen, darunter Wolf, Wildkatze und Biber.[25]

Verantwortlich für die Zusammenstellung des Gesamtberichts und dessen Meldung an das Münchener Finanzministerium und letztlich an Wagner war in Zwiesel der Forstmeister Klein. Wie ein Vergleich der einzelnen Meldungen an Wagner mit seinem Gesamtbericht an das Ministerium zeigt, verzichtete Klein im Gegensatz zu anderen Forstämtern[26] und auch unter Missachtung des Wunsches der Münchener Zentrale, solche in Parenthese zu nennen, auf die Weitergabe von detaillierteren Ortsangaben, wie sie die Förster Zwiesler Waldhaus, Hohenbogen und Weißenstein vorgenommen hatten. Klein beschränkt sich auf die Nennung von Häufigkeiten sowie die grobe, typisierende Beschreibung von Habitaten: Ortschaften, Waldungen, Flüsse und Bäche. Damit ist sein Bericht im Verhältnis aller 119 bayerischen Forstämter recht dünn; auch quantitativ ist er mit 98 Wörtern überschaubar und einer der kürzesten (die Berichte sind im Durchschnitt 295 Worte lang.[27]

---

[25] https://www.nationalpark-bayerischer-wald.bayern.de/natur/tiere/index.htm

[26] Vgl. den Gesamtdatensatz auf Zenodo, wie Anm. 17.

[27] Vgl. auch Malte Rehbein, „From Historical Archives to Algorithms: Reconstructing Biodiversity Patterns in 19th Century Bavaria". Diversity 17, Nr. 5 (26. April 2025): 315. https://doi.org/10.3390/d17050315 und Rehbein, Malte. „Taciturn Franconia? – Some Statistical Exploration of AOD1845". Computational Historical Ecology (blog), 1. November 2024. https://che.hypotheses.org/198. Da der Gesamtbericht des Forstmeisters so wortkarg ist, lohnt sich der Blick in die Zuarbeiten seiner Revier- und Forsteiförster umso mehr.

Auffällig ist auch, dass Forstmeister Klein das aus seinem Revier Bodenmais gemeldete Vorkommen des Luchses (*Lynx lynx*) nicht nach München weitergereicht hat. Heute erfährt gerade die Luchspopulation im Bayerischen Wald große Aufmerksamkeit: Seit den 1970er Jahren erfolgte, auf Initiative des damaligen Vorsitzenden von „BUND Naturschutz" Hubert Weinzierl, zusammen mit Bernhard Grzimek (1909-1987),[28] die Wiederansiedlung von Exemplaren aus dem seiner Zeit noch durch den Eisernen Vorhang getrennten, angrenzenden tschechischen Šumava. Man konstatierte, dass das Tier seit 150 Jahren „aus unseren Wäldern verschwunden" war;[29] so kann der Bericht des Revierförsters von Asch aus dem Jahr 1845 über „wenige" Luchse als glaubwürdig betrachtet werden und zugleich die gegenwärtigen Bemühungen zur Wiederansiedlung und Bestandssicherung zusätzlich informieren.[30]

Weiten wir den Blick: In seinem Bericht an die Königlich Bayerische Akademie der Wissenschaften stellt Andreas Wagner nach Auswertung der Forstamtsmeldungen sowie weiterer Quellen 1846 zunächst fest, dass „der Luchs (*Felix Lynx*) [...]" in den Jahren zwischen 1828 und 1846 „[...] im bayerischen Hochgebirge [...] wohl ganz verschwunden ist, und in demselben nur noch zuweilen als größte Seltenheit auf seinen Streifereien aus Tyrol her verspürt wird. Ein solches Resultat ergiebt sich aus den Berichten der k[öniglichen] Forstämter [...]" (652/653). Hierzu zählen die Berichte der Salforste Rosenheim und Tegernsee sowie der oberbayerischen Forstämter Benediktbeuern und Partenkirchen. Wagner weiter: „In den andern Theilen des Landes [außer dem Hochgebirge] sind die Luchse seit undenklichen Zeiten verschwunden. Vor zweyhundert Jahren schoß man im Fichtelgebirge noch manchmal einen Luchs, seitdem weiß man daselbst nichts mehr von ihnen. Vom bayerischen Walde giebt Schrank nach den Mittheilungen Poschinger's 1798 an,[31] daß der Luchs alle Winter aus Böhmen in die Waldungen des Landgerichts Zwiesel herüberstreife; die Berichte der k[öniglichen] Forstämter Schönberg, Wolfstein und Zwiesel geben einfach an, daß er im bayerischen Wald nicht vorkomme" (653/654). Es bleibt zu spekulieren, wie Wagners Bericht ausgefallen wäre, hätte Forstmeister Klein die Meldung aus dem ihm untergeordneten Revier Bodenmais weitergereicht. Warum Klein dies unterließ, obwohl er es im gleichlautenden Fall des Uhus (*Bubo bubo*) sehr wohl tat, muss unklar bleiben. Wagner jedenfalls schließt seinen Bericht mit einer Vermutung über das Verschwinden dieses Tieres in Bayern: „Die schnelle Vertilgung der Luchse im bayerischen Hochgebirge ist hauptsächlich der von der k[öniglichen] Regierung früher ausgesetzten hohen Prämie zuzuschreiben, indem für jedes eingelieferte Stück 75 Gulden bezahlt wurde. Mit der Verminderung dieser Raubthiere hat die Vermehrung des übrigen Wildstandes gleichen Schritt gehalten" (653). Auch heute stehen die Fragen des Verhältnisses von menschlicher Jagd, Rotwildbestand und Bestandserhaltung des Luchses auf der Tagesordnung.

---

[28] Auch der zuletzt in Passau wohnhafte Journalist und Autor Horst Stern (1922-2019, u.a. „Sterns Stunde") ist in diesem Zusammenhang zu nennen, vgl. Marco Heurich / Christof Mauch (Hrsgg), Urwald der Bayern: Geschichte, Politik und Natur im Nationalpark Bayerischer Wald, Göttingen 2020.

[29] https://www.bund-naturschutz.de/tiere-in-bayern/luchs

[30] Umfassend zum Luchs in Bayerischem Wald und Šumava siehe: Marco Heurich/ Karl Friedrich Sinner. Der Luchs. Die Rückkehr der Pinselohren, Amberg 2012.

---

[31] Wagner nennt keine weiteren Angaben zu dieser Quelle. Es handelt es sich um den Theologen, vormaligen Jesuiten und Professor für Mathematik und Physik an der Ludwig-Maximilians-Universität Franz de Paula von Schrank (1747-1835), Mitglied der Bayerischen Akademie der Wissenschaften, Direktor des Botanischen Gartens München und Mentor von Martius'. In seiner „Fauna boica. Durchgedachte Geschichte der in Baiern einheimischen und zahmen Thiere" (3 Bde., Nürnberg 1798-1803) schreibt von Schrank zur „Luchskaze" (sic): „Wohnort: in Böhmen; aber er kommt alle Winter in die Waldungen des Landgerichts Zwiesel herüber" (Bd. 1.1 S. 52). Beim Urheber der besagten „Mittheilungen" handelt es sich offenkundig um ein Mitglied des niederbayerischen Adelsgeschlechts der Freiherren von Poschinger. Schrank nennt seinen Informanten „Herr Ign. v. Poschinger" (Schrank: 52), und es liegt nahe, dass wir es mit Ignaz Dominikus Poschinger (1747–1803) zu tun haben, einem Weltpriester, der mit seinen Brüdern 1790, während des Interregnums zwischen Joseph II. und Leopold II. vom Kurfürsten von Pfalzbaiern Karl Theodor in dessen Funktion als Reichsvikar den Reichsadel verliehen bekam sowie, zusätzlich, in die Reichsritterschaft aufgenommen wurde, vgl. Genealogisches Handbuch des Adels, Reihe Freiherrliche Häuser B Bd. V (1970) s. v. „Poschinger v. Frauenau", S. 278-281, hier 278. Ignaz von Poschinger gehörte aber nicht nur dem Reichsadel an, sondern auch der landständischen Aristokratie im Kurfürstentum Pfalzbaiern, vgl. das „Verzeichniß der Staende Niederbaierns, Rentamts Straubing, nach dermaliger Verfaſſung, in: Das Churfuerſtliche Rentamt Straubing, Niederlands Baiern […] Beſchrieben von Nepomuk Felix des H. R. R. Grafen Zech […], o. O. 1795, S. 32-40, hier 39, in welchem Ignaz „wegen Traexlried und Wettzell" als Mitglied des „Ritter= und Adelstandes" „vermoeg [seiner] der Landtafel einverleibten Beſitzungen" (vgl. ebd., S. 32) angeführt ist. Julia Bloemer, Empirie im Mönchsgewand. Naturforschung in süddeutschen Klöstern des 18. Jahrhunderts, Göttingen 2022, S. 78, bezeichnet Ignaz von Poschinger als „Glashüttenmeister", der auch Franz Xaver Epp (1733-1789), einem Jesuiten, der u. a. zur Meteorologie publizierte, als Wetterbeobachter, zunächst in Wald (?), dann von 1786 an in Frauenau gedient habe. Siehe auch: [N.N.] Werner., Epp, Franz Xaver, in: Allgemeine Deutsche Biographie 6 (1877), S. 157-158 [Online-Version], URL: https://www.deutsche-biographie.de/pnd100124259.htm (eingesehen 15. Mai 2025).

Blicken wir auf eine weitere Tierart: Anders als manch andere Forstämter berichtet Zwiesel nicht über historische Vorkommen von Braunbären (Ursus arctos). Hier hilft uns Wagners Bericht, das Bild über die Fauna im Forstamt Zwiesel weiter zu vervollständigen: „Länger haben sich die Bären im bayerischen Walde und zwar als Standwild gehalten, denn Poschinger sah noch, wie Schrank berichtet, zu Ende des vorigen Jahrhunderts im Landgerichte Zwiesel diese Thiere als ständige Bewohner, auch giebt derselbe Beobachter an, daß er einmal im Oktober drey Junge mit ihrer Mutter beysammen gesehen hätte. Wann der Bär daselbst ausgerottet wurde, kann ich nicht angeben, da sich die k[öniglichen] Forstämter Schönberg, Wolfstein und Zwiesel mit der einfachen Bemerkung, daß derselbe bey ihnen nicht vorkomme, begnügen. Indeß hat unsere Sammlung, wie Wagler[32] berichtet, noch im Jahre 1826 einen Bären von Zwiesel erhalten, doch während des Winters, daher er wahrscheinlich nur aus Böhmen herübergestreift war." (651) Im gesamten Bestand der Forstamtserhebung berichten lediglich die Salforste Tegernsee über eine rezente Sichtung eines Braunbärs „wahrscheinlich aus Illyrien versprengt [...] Der Letzte war von 1826-28 theils im hiesigen, theils im angrenzenden Tyroler Gebirge, wo er erlegt wurde." Ähnlich berichtet das Forstamt Ruhpolding: „Ursus Arctos, wurden im Amts-Bezirke 2 erlegt, und zwar im J: 1822 et 1835, nun nie mehr wahrgenommen"; Hohenschwangau (Kreis Schwaben und Neuburg) meldet die letzte Sichtung im Jahre 1629 und Selb (Oberfranken) für 1710. So erscheint der Bericht aus dem Forstamt Passau bemerkenswert und die Wagner'sche Zusammenfassung bestätigend: „Nicht mehr im bayr. Wald; alte Leute im Walde erinnern sich noch an dessen Vorkommen."

Auch wissenschaftshistorisch sind an dieser Stelle die Ausführungen Ignaz von Poschingers interessant, die Schrank wiedergibt: „In eben diesen Wäldern (des Landgerichts Zwiesel) giebt es sowohl schwarze als braune Bären [...] Der Bär wird gross und schwer; man hat bey uns (in der Hofmark Frauenau), jedoch selten, Stücke zu 4 Centnern erlegt". Auch Schrank selbst differenziert zwischen einem schwarzen Bären, den er auch „Grasbär" und „Ameisenbär" (sic) sowie wissenschaftlich *Ursus niger* (Gmelin 1788) nennt, und einem braunen Bären, den er als „Honigbär" (*Ursus badius*) oder „Pferdebär" (*Ursus suscus*) bezeichnet. In der modernen Taxonomie gibt es eine solche Unterscheidung nicht mehr. Ein Schwarzbär (*Ursus americanus*) wird ausschließlich in Nordamerika verortet; die Taxa *Ursus niger* sowie *Ursus badius* werden heute als Synonyme für *Ursus arctos* betrachtet, dabei wird *Ursus badius* in der historischen Taxonomie gar einzig auf unsere Quelle, Schrank 1798, zurückgeführt (https://www.gbif.org/species/11986608 und https://www.gbif.org/species/221351447). Schrank führt weiter aus: „Herr Zimmermann[33] hält beyde Arten für blosse Varietäten, wie er diess auch von den verschiedenen Fuchsarten, die über den Erdboden verbreitet sind, zu glauben geneigt ist. Ich bin anderer Meynung. Varietäten, Abartungen, Ausartungen müssen eine Ursache haben; diese kann individuelle oder allgemeine Ursache seyn. Ist sie das erste, so müssen die abgearteten Thiere (ich rede bloss von wilden Thieren) selten seyn, z. B. weisse Raben, ganz weisse Schwalben, schwarze Stieglize, u.f.f. Ist sie allgemein, so muss sie alle, oder fast alle Individuen (nämlich nur die stärksten ausgenommen) ebenderselben Art und ebendesselben Landstriches treffen. Was die Schnauze des einen Bären verlängert, muss auch die des andern verlängern, der mit ihm wohnt, und alle Umstände theilt" (57/58). So führt uns der wortkarge Bericht aus dem Zwieseler Forstamt in eine wissenschaftlich spannende Diskussion um die Evolutionsbiologie, die sich mit Alfred Russel Wallace und Charles Darwin und der Veröffentlichung von "The Origin of Species" 1859 erst noch entwickeln sollte. Andreas Wagner lehnte übrigens die Evolutionstheorie leidenschaftlich ab.

---

[32] Johann Georg Wagler, Einzelne Beyträge zur bayerischen Fauna. Säugethiere und Vögel, in Isis. Encyclopädische Zeitschrift, vorzügl. für Naturgeschichte, vergleichende Anatomie u. Physiologie, Nr. 11 (1828), 1140–44.

Abgesehen von den Fischen (siehe unten) geht Wagner in seinem Bericht an die Bayerische

---

[33] Eberhard August Wilhelm von Zimmermann (1743-1815, GND 11882435X). Bei Schranks Verweis auf „Geogr. Gesch. I. 210" handelt es sich um: \zimmermannGeographischeGeschichteMenschen1778. Der X. Abschnitt (S. 209-213) behandelt den Bären.

Akademie noch ein drittes Mal auf Zwiesel ein. So notiert er über Auerhuhn (*Tetrao urogallus*), Haselhuhn (*Tetrastes bonasia*) und Birkhuhn (*Tetrao tetrix*): „Im bayerischen Walde (Forstamt Wolfstein, Schönberg und Zwiesel) kommen Auerhühner und Haselhühner zum Theil ziemlich häufig vor; das Birkhuhn dagegen ist in den beyden ersten Bezirken gar nicht und im letztern nur sehr selten vorhanden" (669). Es sind auch diese drei Vogelarten, die Wagner in seiner Übersichtskarte neben wenigen anderen Tierarten im Raum Zwiesel darstellt. Die Einzelberichte aus den Revieren erlauben uns nun aber eine detailliertere biogeographische Kartierung, die wir in Abb. 5 in einer Gesamtschau sowie — exemplarisch für eine Tierart — mit Blick auf das Auerhuhn in Abb. 6 präsentieren. Die hierfür verwendeten Daten sind in Tab. 3 (Säugetiere, Vögel) und Tab. 4 (Fische) zusammengefasst. Für die Kartierung haben wir die Angabe „Hochwaldungen" pauschal als Höhenlagen über 900m (Gr. Arber), 800m (Osser, Keitersber, Hohenbogen), 600m (Haidstein), um den jeweiligen Gipfel herum, angenommen.

| Revier | Toponym | Tierart | | Häufigkeit |
|---|---|---|---|---|
| Zwieseler Waldhaus | Ludwigsthal | Dachs | *Meles meles* | selten |
| | | Steinmarder | *Martes foina* | nicht sehr häufig |
| | Lindberg | Dachs | *Meles meles* | selten |
| | | Steinmarder | *Martes foina* | nicht sehr häufig |
| Hohenbogen | Hochwaldungen der Ossa | Edelmarder | *Martes martes* | häufig |
| | | Auerhuhn | *Tetrao urogallus* | nicht selten |
| | | Haselhuhn | *Tetrastes bonasia* | sehr häufig |
| | | Schnepfe | *Scolopax rusticola* | sehr häufig |
| | Hochwaldungen Hohenbogen | Edelmarder | *Martes martes* | häufig |
| | | Reh | *Capreolus capreolus* | häufig |
| | | Auerhuhn | *Tetrao urogallus* | nicht selten |
| | | Haselhuhn | *Tetrastes bonasia* | sehr häufig |
| | | Schnepfe | *Scolopax rusticola* | sehr häufig |
| | Hochwaldungen Keitersberg | Edelmarder | *Martes martes* | häufig |
| | | Auerhuhn | *Tetrao urogallus* | nicht selten |
| | | Haselhuhn | *Tetrastes bonasia* | sehr häufig |
| | | Schnepfe | *Scolopax rusticola* | sehr häufig |
| | schwarzer Regen und Nebenbäche | Fischotter | *Lutra lutra* | nicht selten |
| | weißer Regen und Nebenbäche | Fischotter | *Lutra lutra* | nicht selten |
| | Hochwaldungen Haidstein | Reh | *Capreolus* | häufig |
| | | | *capreolus* | |
| | | Haselhuhn | *Tetrastes bonasia* | sehr häufig |
| | | Schnepfe | *Scolopax rusticola* | sehr häufig |
| | Hochwaldungen Arber | Auerhuhn | *Tetrao urogallus* | nicht selten |
| | | Haselhuhn | *Tetrastes bonasia* | sehr häufig |
| | | Schnepfe | *Scolopax rusticola* | sehr häufig |
| | Privatwald Hollerberg | Auerhuhn | *Tetrao urogallus* | geringe Anzahl |
| Weißenstein | | Haselhuhn | *Tetrastes bonasia* | häufiger |
| | Privatholz Baiken | Auerhuhn | *Tetrao urogallus* | geringe Anzahl |
| | | Haselhuhn | *Tetrastes bonasia* | häufiger |
| | Rinchnamünderhölzer | Auerhuhn | *Tetrao urogallus* | geringe Anzahl |
| | | Haselhuhn | *Tetrastes bonasia* | häufiger |
| | kgl. Walddistrikt Hollerberg | Auerhuhn | *Tetrao urogallus* | geringe Anzahl |
| | | Haselhuhn | *Tetrastes bonasia* | häufiger |
| | Weissensteinerau | Auerhuhn | *Tetrao urogallus* | geringe Anzahl |
| | | Haselhuhn | *Tetrastes bonasia* | häufiger |
| | Bachwiesen bei Reichnach | Bekassine | *Gallinago gallinago* | strichweise in geringer Anzahl |
| | Richnacher Ohe | Stockente | *Anas platyrhynchos* | ein bis zwei Brutpaare jährlich |

*Tabelle 3: Übersicht der Tierarten, deren Vorkommen mit präziseren (namentlichen) Ortsangaben versehen sind. Der Bericht des Reviers Zwiesel enthält keine genauen Ortsangaben.*

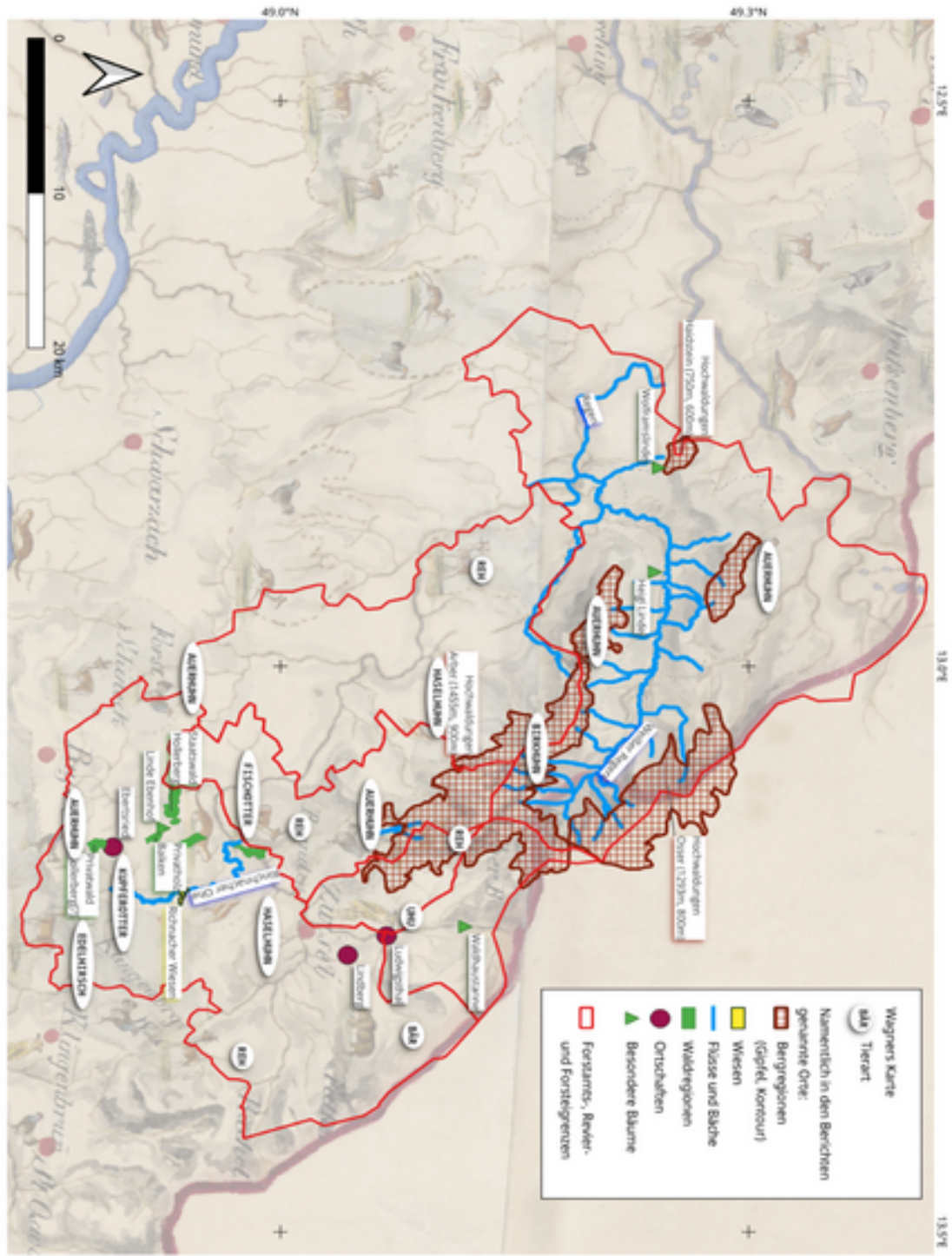

*Abbildung 4. Kartierung der in den Berichten genannten Orte und Regionen, projiziert auf die Wagner'sche Karte von 1846 nebst Identifizierung der dort gezeichneten Tierarten.[34]*

---

[34] Wir danken Kilian Baumgartner für die Geocodierung der Wagnerschen Tierzeichnungen.

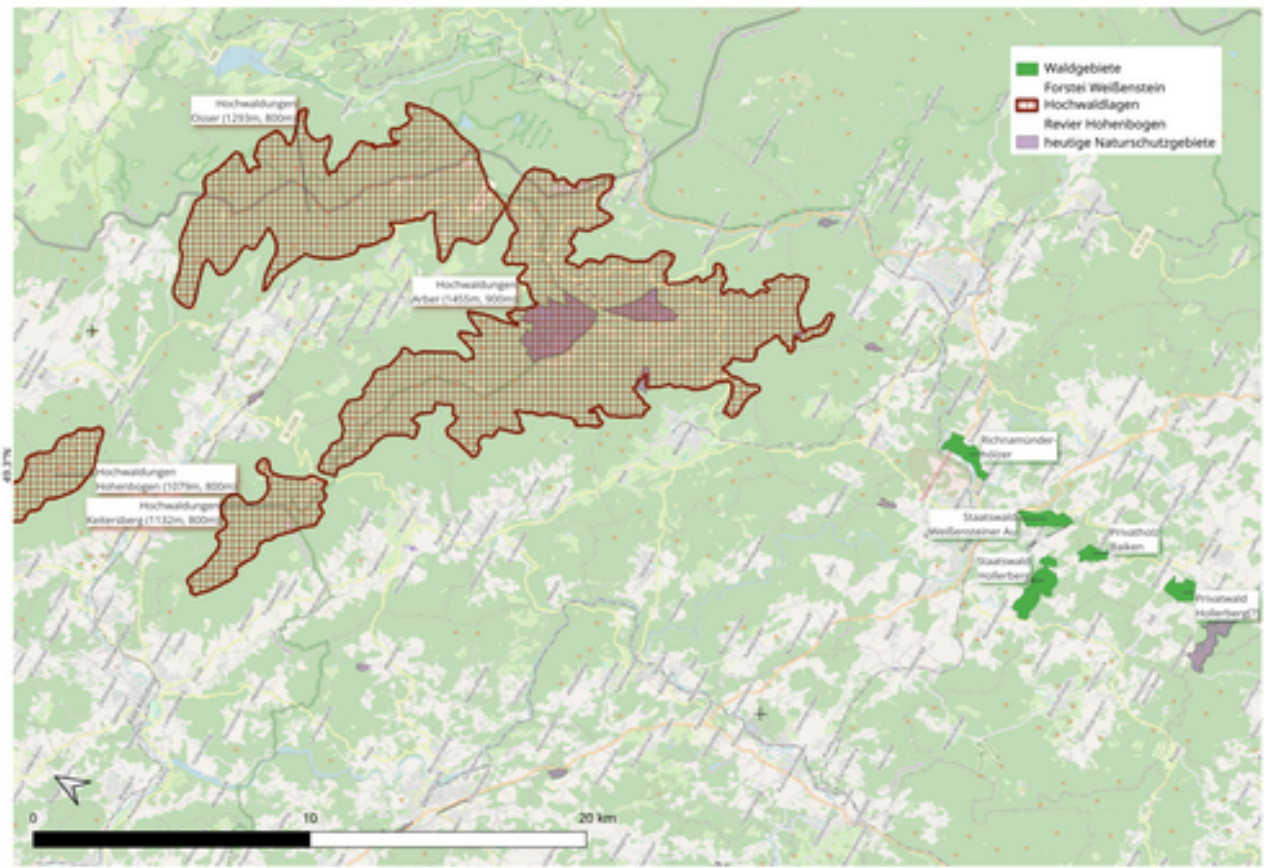

*Abbildung 5. Approximiertes Vorkommen des Auerhuhns (*Tetrao urogallus*) in den namentlich genannten Gebieten des Reviers Hohenbogen und der Forstei Weißenstein. Außerdem dargestellt sind die heutigen Naturschutzgebiete. Quellen: Bundesamt für Naturschutz, OpenStreetMap. Man beachte die Rotation der Karte.*

Wenden wir uns abschließend der letzten Zeile im von Wagner angelegten Formular zu den Tiervorkommen zu. Sie dient der Erhebung „von Fischen die merkwürdigsten", wobei wir „merkwürdig" hier im Sinne von bemerkenswert (engl: *remarkable*), nicht aber seltsam (*strange*) zu verstehen haben. Auch hier ist auffällig, dass Forstmeister Klein die doch deutlich differenzierteren Berichte seiner Revierförster in diesem Detaillierungsgrade nicht weitergibt. Wieder werden die Ortsangaben nicht übernommen, und es fehlt auch die Nennung des Vorkommens des Aals (*Anguilla anguilla*) — eine Art, die heute in Deutschland als stark gefährdet gilt[35] und für die in der *Global Biodiversity Information Facility (GBIF)*

---

[35] https://www.rote-liste-zentrum.de/de/Detailseite.html?species_uuid=6a1a9828-ca7b-422b-82b6-

keine Sichtung in Bayern nordöstlich der Donau verzeichnet ist.[36]

| Revier | Toponym | Tierart | | Häufigkeit |
|---|---|---|---|---|
| Hohenbogen | schwarzer Regen | Huchen | *Hucho hucho* | k.A. |
| | | Schleie (vermutl.) | *Tinca tinca* | k.A. |
| | | Äsche | *Thymallus thymallus* | k.A. |
| | | Hecht | *Esox lucius* | k.A. |
| | weißer Regen | Forelle | *Salmo trutta fario* | k.A. |
| | Gebirgsbäche des Arber | Forelle | *Salmo trutta fario* | sehr häufig |
| | Gebirgsbäche des Osser | Forelle | *Salmo trutta fario* | sehr häufig |
| | Gebirgsbäche des Keitersberg | Forelle | *Salmo trutta fario* | sehr häufig |
| Bodenmais | Regen | Äsche | *Thymallus thymallus* | nicht zu häufig |
| | | Huchen | *Hucho hucho* | nicht zu häufig |
| Weißenstein | Richnachbach | Forelle | *Salmo trutta fario* | k.A. |
| | Kühbach | Forelle | *Salmo trutta fario* | k.A. |
| | Sengbächl(?) | Forelle | *Salmo trutta fario* | k.A. |
| | Haidbächl | Forelle | *Salmo trutta fario* | k.A. |
| | Schloßauer Ohe | Forelle | *Salmo trutta fario* | k.A. |
| | | Äsche | *Thymallus thymallus* | k.A. |
| | | Hecht | *Esox lucius* | k.A. |
| | | Huchen | *Hucho hucho* | k.A. |
| | Regen | Äsche | *Thymallus thymallus* | k.A. |
| | | Hecht | *Esox lucius* | k.A. |
| | | Huchen | *Hucho hucho* | k.A. |
| | Richnacher Ohe | Äsche | *Thymallus thymallus* | k.A. |
| | | Hecht | *Esox lucius* | k.A. |

*Tabelle 4. Übersicht der Fischarten, deren Vorkommen mit präziseren (namentlichen) Ortsangaben versehen sind. Die Berichte der Reviere Zwiesler Waldhaus und Zwiesel enthalten keine genauen Ortsangaben. k.A. = keine Angabe.*

Auch die in allen Revieren genannte Äsche (*Thymallus thymallus*) gilt heute als stark gefährdet. Sie kam seinerzeit im Regen sowie in weiteren größeren Bächen des Forstamts nicht selten vor. Ein mäßiges Vorkommen des Huchen (*Hucho hucho*), heute ebenfalls stark

---

87b6cda71441&species_organismGroup=Meeresfische%20und%20-neunaugen&q=aal

[36] Abfrage https://www.gbif.org/occurrence/gallery?
has_coordinate=true&has_geospatial_issue=false&taxon_key=5212973&geometry=POLYGON((12.64266%2049.429
21,12.39942%2048.99667,12.77779%2048.89461,12.994%2048.82793,13.18318%2048.68539,13.36561%2048.6095,1
3.48722%2048.6229,13.54803%2048.92125,13.39263%2048.95232,13.4129%2049.08525,12.64266%2049.42921))
<8. Mai 2025>.

gefährdet, wird vor allem im Regen gemeldet. Während der Hecht (*Esox lucius*) als ungefährdet gilt, mag das detailliert beschriebene weite Vorkommen der Steinforelle (heute: kleinwüchsige Bachforellen, *Salmo trutta fario*) von besonderem Interesse für heutige ökologische Betrachtungen sein, gilt die Forelle inzwischen doch als gefährdet und als ein „deutliches Warnsignal für größere, klimabedingte Biodiversitätsveränderungen in Fließgewässern."[37]

Die Berichte über die Fischvorkommen illustrieren auch Herausforderungen in der historischen Ökologie, wie sie aus der Quellensprache resultieren. So ist eine weitere Auffälligkeit die Nennung einer „Schiele", die Klein so aus dem Bericht Hohenbogen übernommen haben könnte. Einen Fisch mit diesem Namen gibt es jedoch nicht, auch eine historische oder dialektale Schreibweise konnten wir nicht finden. Vermutlich ist die weit verbreitete Schleie (*Tinca tinca*) gemeint. Die von Klein in Parenthese zugefügte Bezeichnung „Hauchforelle" für den Huchen konnte hingegen nachgewiesen werden.[38]

Insgesamt wirkt der an das Ministerium übermittelte Bericht des Forstmeisters Klein aber durch die starken Verkürzungen recht wenig an dem ministeriell gestellten Auftrag interessiert und durch Fehler wie der falschen Bezeichnung einer Fischart als nicht übermäßig aussagekräftig. Es mangelte Klein dabei sicher nicht an fachkundigen Zuarbeiten seiner ihm unterstellten Revierförster, die die Erstellung eines umfassenderen Berichts wie sie andere Forstämter Wagner vorlegten, ermöglicht hätten. Wagner schließlich stellte zu den Fischvorkommen zusammenfassend und durch andere Quellen ergänzt 1846 dar:

„Dagegen hat das Donaugebiet nicht wenig andere Arten auszuweisen, die dem Maingebiete ganz abgehen; darunter sind die werthvollsten der Welsoder Waller (*Silurus*

---

[37] https://www.rote-liste-zentrum.de/de/Detailseite.html?species_uuid=6a0ba544-5904-4d98-b48b-bd917dae1f43&species_organismGroup=S%C3%BC%C3%9Fwasserfische%20und%20Neunaugen&q=Salmo%20trutta

[38] Deutsches Wörterbuch von Jacob und Wilhelm Grimm. Lfg. 3 (1870), Bd. IV,II (1877), Sp. 572, Z. 20 (gefunden über das Digitale Wörterbuch der deutschen Sprache: https://www.dwds.de).

*glanis*) und aus der Salmgattung der Huche (*Salmo hucho*), der Salbling [sic] (*Salmo salvelinus*),[39] der Silberlachs (*Salmo lacustris*) und der Renke (*Salmo s. Coregonus Wartmanni*). Unter den genannten hat der Huche (in Schwaben Rothfisch genannt) die größte Verbreitung und nächst ihm der Wels; ersterer kommt in der ganzen Donau von Günzburg bis Passau und in allen ihren südlichen Zuflüssen vor, und ist unter ihnen, meines Wissens, der einzige der eben verzeichneten, der auch noch in den nördlichen Zuflüssen, aber nur in denen des bayerischen Waldes (Forstämter Wolfstein, Schönberg und Zwiesel) aufsteigt. Alle übrigen sind unter den Fischen der Naab, Vils, Altmühl und den andern nördlichen Seitenflüssen der Donau nicht mit aufgezählt, daher auf den Hauptstrom und seine südlichen Gewässer beschränkt. Von diesen Donaufischen kommt nur der Wels nordwärts noch in andern deutschen Flüssen vor, während der Salbling nicht eher als in den Seen von Cumberland und Westmoreland wieder gefunden wird." (680)

## Baumarten im Forstamt Zwiesel: Schema B

Im Fragebogen Schema B über die Bäume werden nicht etwa einzelne Baumarten abgefragt, handelt es sich hierbei um einen Fragenkatalog mit offenen Fragen:

1. Welche Baumarten im Amtsbezirk bilden die größern reinen und welche die größern gemischten Bestände?
2. Welche Bäume kommen außer geschlossenen Beständen auf lichten Plätzen, Viehweiden, Gemeinde-Gründen u. s. w. vor, oder werden daselbst und in den Umfriedungen der Grundstücke gezogen?
3. Kommen seltenere Baumarten, wie z. B. im Gebirge, die Lerche, die Zirbelkiefer, Eibe; im Flachlande Pyrus torminalis[40], Sorbus domestica u. a. vor?
4. Findet sich Quercus pedunculata oder Robur[41], oder vielleicht beide und beiläufig in welchem Verhältnisse?

---

[39] Linnaeus, 1758. Synonym zu *Salvelinus umbla* (Seesaibling oder Rotforelle).

[40] Wohl Elsbeere (*Sorbus torminalis*).

[41] *Quercus pedunculata* und *Quercus robur* werden heute beide als Synonyme für die Stieleiche gebraucht.

5. Ist tilia parvifolia oder grandifolia[42] an den Orten, wo gewöhnlich Linden gepflegt werden, im Gebrauche, und wie gedeihen sie?
6. Welche Sträucher sind auf vorkommenden Haiden oder Moorgründen vorherrschend?
7. Finden sich einzeln durch ihre Dimensionen und ihr Alter besonders ausgezeichnete Individuen von Baumarten? Man wünscht hiebei außer der Bestimmung der Spezies möglichst genaue Angaben des Standortes, der Dimensionen, des muthmaßlichen Alters und der allenfalls über solche Individuen gängigen Traditionen und Sagen?

Die Antworten auf die ersten sechs Fragen zur Baumverbreitung und -zusammensetzung im Forstamt geben einen detaillierten Einblick in die damalige Vegetation und Nutzung der Waldlandschaft. Die im Zuge der historischen Erhebung genannten Baumarten spiegeln eine bemerkenswerte Vielfalt wider und haben teils spezifische Standortansprüche. So sind Arten wie Lärche, Zirbelkiefer und Eibe typische Vertreter höherer Gebirgslagen, wo sie natürlicherweise vorkommen. Heute gelten sie jedoch als selten, da sie auf besondere Umweltbedingungen angewiesen sind und historisch stark genutzt wurden.[43] Auch im Flachland wurden bemerkenswerte Baumarten genannt, etwa *Sorbus torminalis* (Elsbeere) und *Sorbus domestica* (Speierling), die aufgrund ihres ökologischen Werts Bedeutung erlangten und geschätzt wurden.[44]

Das Forstamt Zwiesel liefert in seinen Antworten weitere Einblicke in die damalige Vegetation und Nutzung. Demnach bildeten Fichte, Föhre (Kiefer), Rotbuche und Birke regional größere reine Bestände. In gemischten Beständen traten vor allem Fichte, Tanne und Buche auf. Edlere Holzarten wie Ahorn, Esche und Ulme ergänzten diese Zusammensetzung gelegentlich, während unedlere Arten wie Birke, Aspe, Erle und Föhre

---

[42] *Tilia parvifolia Erhr.* und *Tilia grandifolia* Ehrh. (1790) sind heute eher ungebräuchlich und Synonyme für die Winterlinde (*Tilia cordata*) und die Sommerlinde (*Tilia platyphyllos*).

[43] Auf ein vermindertes Vorkommen der Eibe bereits in der Mitte des 19. Jahrhunderts deutet der Bericht von Karl Frhr. v. Asch, Revierförster von Bodenmais hin: „Kommt nur die Lerche vor, und zwar sehr wenig, von Eiben findet sich keine Spur mehr." (StALa, Forstamt Zwiesel A 988, S. 11.)

[44] Franz, Christine; Müller-Kroehling, Stefan: Elsbeere und Speierling in Bayern. Bemühungen um ihren Erhalt, Anbau, Waldbau und Holzverwertung (Corminaria 12, 1999), S. 7.

vorwiegend auf offenen Flächen oder weniger genutzten Standorten vorkamen.

Auch zu Beginn des 20. Jahrhunderts blieb die Fichte die dominierende Baumart im Bayerischen Wald. L. Leythäuser, Forstrat in Landshut, beschreibt 1906, dass die Fichte, aufgrund ihrer Engringigkeit und Spaltbarkeit, weiterhin die häufigste Holzart war, der die Tanne, Buche, Ahorn, Birke und Föhre folgten, die somit für den gesamten Bayerischen Wald prägende Baumarten blieben.[45] Solche Erhebungen dokumentieren nicht nur die historische Verbreitung von Baumarten, sondern geben auch Hinweise auf ihre ökologischen und ökonomischen Funktionen. *Quercus robur* (Stieleiche) war dabei ökologisch besonders bedeutsam, da sie zur Entstehung lichter Wälder beitrug, die eine hohe Artenvielfalt förderten. Gleichzeitig könnte sie wirtschaftlich von Interesse gewesen sein, insbesondere im Hinblick auf die Holznutzung.[46] Ihre Relevanz spiegelt sich auch in dem fünf Jahre später erstellten, vierseitigen Fragebogen „Bemerkungen über das Vorkommen nachgenannter Holzgewächse" wider, in dem sie neben 46 weiteren Baumarten explizit aufgeführt wurde.[47] Vergleichbare Überlegungen könnten auch für *Tilia cordata* (Winterlinde) und *Tilia platyphyllos* (Sommerlinde) gegolten haben, die kulturhistorisch und als Insektenweide von Bedeutung waren. Insgesamt spiegeln die Antworten des Forstamts Zwiesel sowohl die ökologische Vielfalt als auch die damaligen Nutzungsinteressen wider, die sich von reinen Wirtschaftswäldern bis hin zu biodiversen Mischbeständen erstreckten.

Abschließend wollen wir die siebte Frage näher betrachten, die uns Einblicke in die damalige Wahrnehmung von Natur bietet: „Finden sich einzeln durch ihre Dimensionen und ihr Alter besonders ausgezeichnete Individuen von Baumarten? Man wünscht hiebei außer der Bestimmung der Spezies möglichst genaue Angaben des Standortes, der Dimensionen,

---

[45] Leythäuser, L.: Wirtschaftliche und industrielle Rundschau. Im Gebiet des Inneren Bayerischen Waldes, Passau 1906, S. 10.

[46] Auch heute ist die Stieleiche von großer Wichtigkeit im Hinblick auf Ökologie und die Holznutzung: https://forest.jrc.ec.europa.eu/en/european-atlas/qr-trees/pedunculate-oak/

[47] StALa, Forstamt Zwiesel A 988, S. 51-54.

des muthmaßlichen Alters und der allenfalls über solche Individuen gängigen Traditionen und Sagen?"[48]

Während die Forstreviere Bodenmais und Zwiesel angaben, dass solche bemerkenswerten Bäume in ihrem Bereich nicht vorkämen, lieferten die Reviere Zwiesler Waldhaus, Hohenbogen und Weißenstein detaillierte Beschreibungen ihrer ältesten Baumexemplare:

Im Revier Zwieslerwaldhaus sollen mehrere Tannen und Fichten mit beeindruckenden Dimensionen existieren, darunter Exemplare mit einem Stammdurchmesser von etwa 2 Metern und einer Höhe von 50 bis 55 Metern. Einige dieser Bäume waren im Jahr 1845 bereits weit über 300 Jahre alt. Der Revierförster hob eine Tanne besonders hervor, die sich etwa eine halbe Stunde Fußmarsch von Zwieslerwaldhaus entfernt befindet. Diese soll einen Stammdurchmesser von etwa 2,2 Metern und ein geschätztes Alter von rund 300 Jahren gehabt haben. Vermutlich handelt es sich hierbei um die heute als „Urwaldtanne" bekannte Tanne im Nationalpark Bayerischer Wald.[49] Die Beschreibung der Entfernung zur „Einöde Zwieslerwaldhaus" passt zu ihrem Standort, und der Baum, heute auf ein Alter von etwa 600 Jahren geschätzt, beeindruckte wohl bereits 1845 durch seine Größe und sein hohes Alter. Auch wenn die Altersangabe des Försters abweicht, könnte es sich dennoch um denselben Baum handeln, da das geschätzte Alter auf Vergleichen mit einer benachbarten Tanne basiert, die in den 1960er Jahren bei einem Sturm umgefallen war und man damit die Jahresringe zählen und von ihr auf die „Urwaldtanne" hochrechnen konnte.[50]

Im Revier Hohenbogen wurden zahlreiche Buchen, Tannen und Fichten mit einem geschätzten Alter von 200 bis 300 Jahren erwähnt. Besonders bemerkenswert waren für den Revierförster jedoch zwei Linden in den Ortschaften Ried und Gotzendorf. Eine davon,

---

[48] StALa, Forstamt Zwiesel 988, S. 27.

[49] 49.099N, 13.230E EPSG:4326 WGS 84. Roloff, Andreas: Weiß-Tanne bei Bayerisch Eisenstein im Nationalpark Bayerischer Wald (Landkreis Regen im Regierungsbezirk Niederbayern, https://nationalerbe-baeume.de/project/weiss-tanne-bei-bayerisch-eisenstein-im-nationalpark-bayerischer-wald-landkreis-regen-im-regierungsbezirk-niederbayern/.

[50] Ebd.

„die Stärkere", wurde wie folgt beschrieben: Sie hat einen Stockdurchmesser von etwa 4,7 Metern und hat ein geschätztes Alter von rund 300 Jahren. Ihr Inneres ist bis auf eine Höhe von 4,4 Metern hohl und diente als Kapelle, die Krone weit verzweigt und gesund.

Diese Beschreibung lässt vermuten, dass es sich um die heute als Wolframslinde bekannte Linde in Ried handeln könnte.[51] Dieser Baum, dessen Krone auch heute noch gesund und blühend ist, ist ebenfalls innen vollständig hohl. Das Alter des Baumes wird allerdings heute auf etwa 1000 Jahre geschätzt, was die damalige Angabe von 300 Jahren deutlich übertrifft.[52] Von Hans-Joachim Fröhlich wird die Wolframslinde zu den „Baumveteranen in Bayern" gezählt: In seinem Beitrag zum 250. Jubiläum der Bayerischen Forstverwaltung zweifelt er jedoch das Alter der Linde an, denn es sei „geradezu erstaunlich, [...] wie magisch die Zahl 1.000 in Verbindung zu Bäumen tritt, ohne dass man den Beweis dafür antreten kann".[53]

In Gotzendorf gibt es zudem die sogenannte Räuber-Heigl-Linde, die heute auf ein Alter von etwa 400 Jahren geschätzt wird.[54] Diese könnte das in der Erhebung erwähnte Exemplar aus diesem Ort sein.

Das Revier Weißenstein liefert schließlich eine weitere Beschreibung eines außergewöhnlichen Baumes: „eine Tilia grandifolia von 7 Fuß Stockdurchmesser" die im Jahr 1845 älter als 300 Jahre alt und noch vollkommen gesund gewesen sein soll. Bei den Recherchen konnte ein Baum identifiziert werden, der auf die Beschreibung zutrifft (Abb. 3).

---

[51] 49.202N, 12.824E EPSG:4326 WGS84.

[52] Tourismusverband Ostbayern e.V.: Wolframslinde, https://www.ostbayern-tourismus.de/attraktionen/wolframslinde

[53] Hans-Joachim Fröhlich, Baumveteranen in Bayern, in: Hans Bleymüller, Egon Gundermann, Roland Beck: 250 Jahre Bayerische Staatsforstverwaltung. Rückblicke, Einblicke, Ausblicke. Mitteilungen aus der Bayerischen Staatsforstverwaltung 51, S. 655-664, hier 659.

[54] 49.200N, 12.916E EPSG:4326 WGS 84. Bayerisches Thermenland: Räuber Heigl Linde, https://www.bayerisches-thermenland.de/attraktion/raeuber-heigl-linde

Ein bemerkenswerter Aspekt in der Überlieferung dieses Akts sind die handschriftlichen Ergänzungen mit Bleistift, die wohl zu einem späteren Zeitpunkt hinzugefügt wurden. Auch die 22 Seiten, die die Erhebung behandeln, wurden nachträglich, ebenfalls mit Bleistift, nummeriert. Die weiteren hinzugefügten Ergänzungen bieten interessante Einblicke: Im ausgefüllten Fragebogen des Reviers Zwiesel wurde bei Frage 5 – „Welche Sträucher sind auf vorkommenden Haiden oder Moorgründen vorherrschend?" – der Begriff „Moorgründe" unterstrichen und mit dem Kommentar versehen, diese seien „sehr mangelhaft"[55]. Außerdem, wurde das Vorkommen des Haselstrauchs kommentiert, vermutlich hat der unbekannte Urheber der Kommentare hier das Adjektiv „oft" angefügt.[56] Zum Vorkommen der Kupferotter im Forstamt Zwiesel heißt es: „1902 Verhältnismäßig häufig; damals noch nicht? darum hat sich eben niemand gekümmert!"[57] Diese Anmerkung legt nahe, dass die Kommentare um das Jahr 1902 entstanden sind, also mehr als ein halbes Jahrhundert nach der ursprünglichen Erhebung. Dies belegt nicht nur den Gebrauch der Akte weit nach dessen Entstehen, sondern gibt gleichzeitig exemplarische Hinweise auf Entwicklungen von Flora und Fauna in diesem Zeitraum.

---

[55] StALa, Forstamt Zwiesel 988, S. 14.

[56] Ebd.

[57]

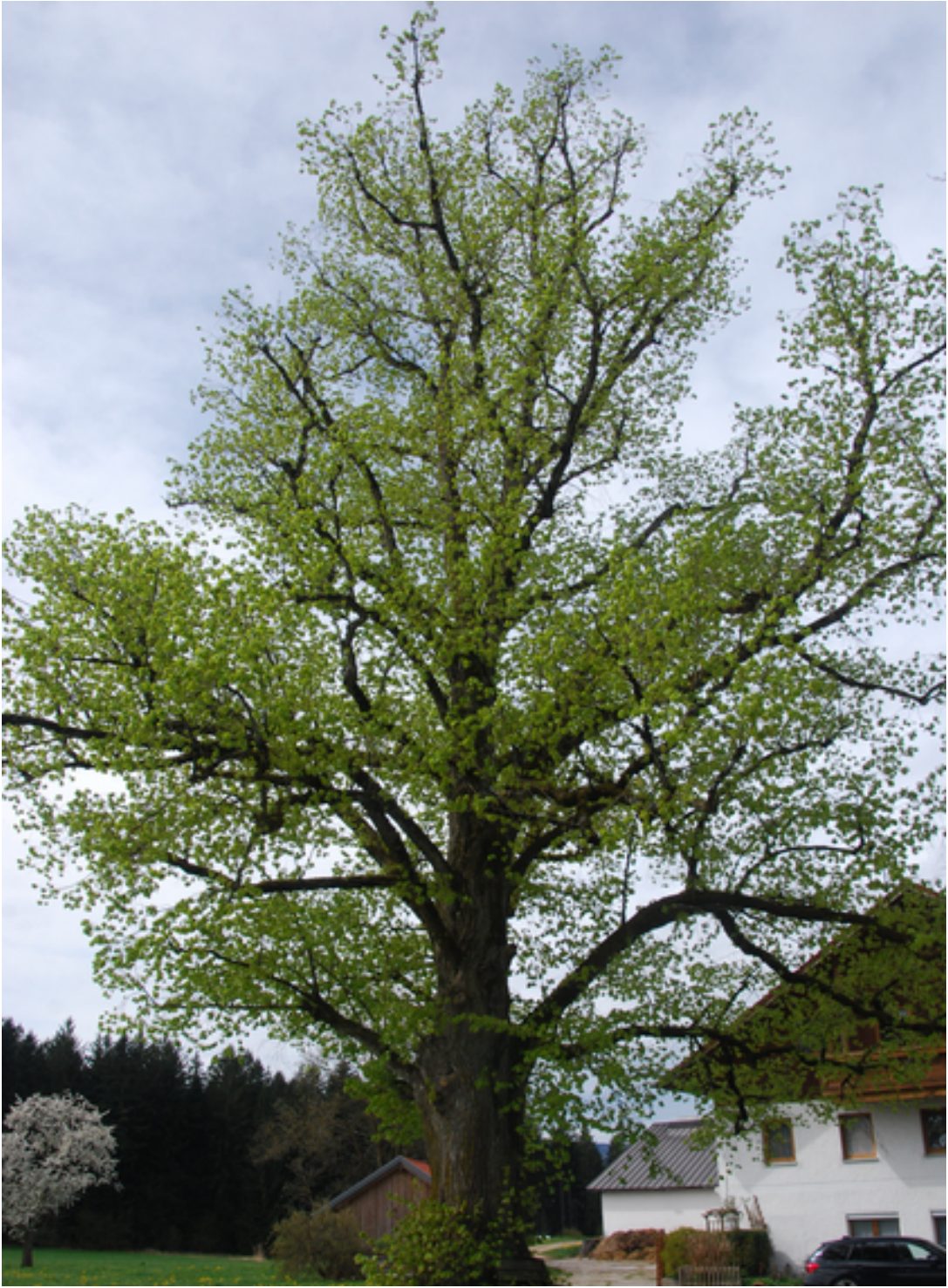

*Abbildung 6. Die 1845 beschriebene Linde bei Ebenhof, Lkr. Regen, 48.937 N, 13.143E, EPSG:4326 WGS 84 (Foto: Rehbein).*

## Ausblick

Trotz der bemerkenswerten neuen Einsichten in die Geschichte der regionalen historischen Ökologie und zur forstlichen Wissensgeschichte des mittleren 19. Jahrhunderts, welche die Datenerhebungen des Forstamtes Zwiesel liefern – insbesondere was das hier ausnahmsweise überlieferte Schema B der bayernweiten Befragung des Jahres 1845 angeht –, bleibt noch viel zu erfahren und zu verstehen: Die Baumerhebung ist auch im Falle Zwiesels nur teilweise überliefert, und weitere Untersuchungen in anderen Archivbeständen mögen zusätzliche Informationen zutage fördern. Insbesondere die Einbindung lokaler Forstamtsüberlieferungen kann helfen, bisher unbeantwortete Fragen zu beantworten.

Das Landshuter Archivale zum Forstamt Zwiesel birgt darüber hinaus noch weitere Erhebungen und Aufzeichnungen. Eben diese sollen, anknüpfend an die bereits unternommenen Forschungen und vorliegenden Publikationen des CHE-Forschungsschwerpunktes am Lehrstuhl für Computational Humanities der Universität Passau, in nächster Zeit u.a. auch im Rahmen einer ebendort betreuten geschichtswissenschaftlichen Masterarbeit weiterverfolgt werden.

Wissenschaften 22 (1846): 649–56, 657-664,665-672,673-680,697-700.

## Literatur

Haas/Gerstmeier/Huter/Rehbein: Von Luchsen und Linden (preprint)Schützen - die Aufgaben der Bayerischen Staasforstverwaltung im Wandel der Zeit". Bayerns Wälder: 250 Jahre, Hefte zur Bayerischen Geschichte und Kultur, 2002.

Tourismusverband Ostbayern e.V. „Wolframslinde", o. J. https://www.ostbayern-tourismus.de/attraktionen/wolframslinde.

Werner., Epp, Franz Xaver, in: Allgemeine Deutsche Biographie 6 (1877), S. 157-158 [Online-Version], URL: https://www.deutsche-biographie.de/pnd100124259.htm (eingesehen 15. Mai 2025).

Wulf, Andrea. The Invention of Nature: How Alexander Von Humboldt Revolutionized Our World, New York, N.Y., 2015; deutsch: Alexander von Humboldt und die Erfindung der Natur, München 2016.
37

# Anhang (Transkription)

[S. 2]

Ad Num. Exh. 16671.                                                                 Landshut am 18ten Aug 1845.
 Einl. Nr. 1783                                                                      Praes[entiert] am 23 Aug. 1845.

<div style="text-align:center">

Im Namen
Seiner Majestät des Königs
von Bayern

</div>

    Das k[öngliche] Forstamt erhaelt hiemit in Folge h[öchster] Fin[anz] Minist[eriel]l[er] Entschliessung vom 13t d[es] M[ona]ts No 12,610 den Auftrag, zur Kenntniß der geographischen Verbreitung der Thiere u. Baumarten in Bayern die beigefügten Schemata A. u. B. mit moeglichster Genauigkeit auszufüllen und solche sodann binnen 14 Tagen in Vorlage zu bringen.—

Kgl. Regierung von Niederbayern
Kammer der Finanzen. -
F[reiherr von] Wulffen[58]
[v]Weinbach[59]

An das k. Forstamt Zwiesel.
    Die geographische Verbreitung der Thier- u. Pflanzarten in Bayern betr[effend]
                                                                                                                                                     Degen[60]

---

[58] Es handelt sich um den damaligen Präsidenten der Kreis-Regierung von Niederbayern (vgl. Hof= und Staatshandbuch des Königreichs Bayern 1845, S. 251), Friedrich Freiherr von Wulffen, zu lesen als „F[reiherr v.] Wulffen". v. Wulffen (1790–1858) wird 1848/49 für den 7. niederbayerischen Wahlkreis Wolfstein als Mitglied in die Frankfurter Nationalversammlung in der dortigen konservativen Fraktion Café Milani einziehen, 1855 wird er zum Präsidenten des Ober=Appellationsgerichts, also der damals (wie heute, als Oberstes Landesgericht) höchsten Jurisdiktionsinstanz im Staate Bayern, avancieren. Zu ihm auch Heinrich Best / Wilhelm Wege, Biographisches Handbuch der Abgeordneten der Frankfurter Nationalversammlung 1848/49 (= Handbuch zur Geschichte des Parlamentarismus und der politischen Parteien, im Auftrag der Kommission für Geschichte des Parlamentarismus und der politischen Parteien hrsgg. von Rudolf Morsey und Gerhard A. Ritter, Bd. 8), Düsseldorf 1996, S. 366 und 392.

[59] Die Lesung der oberen Unterschrift „F Wulffen" wird von dem Sachverhalt untermauert, dass die weiter unten situierte Unterschrift wegen des im Ancien Régime in amtlichen Dokumenten üblichen „Respektsabstands" nur von einem Beamten stammen kann, der in der Regierungshierarchie unter v. Wulffen stand. Ich lese sie als „Weinbach" oder „v Weinbach" – laut Hof= und Staatshandbuch des Königreichs Bayern 1845, l. c., amtierte Ludwig von Weinbach, Ritter des Verdienstordens vom hl. Michael, als „Director" der Kammer der Finanzen der Kgl. Regierung von Niederbayern.

[60] Joh[ann] Phil[ipp] Heinr[ich] Degen war Beamter im „Secretariat beider Kammern", d. h. der Kammer des Innern und der Kammer der Finanzen in der Regierung von Niederbayern, vgl. Hof= und Staatshandbuch des Königreichs Bayern

[Bericht des Revierförsters zu Zwiesler-Waldhaus Frz. Xaver Seninger, 1845, August 29]

[S. 3]

A. Verzeichnis der im Revierbezirk Zwiesler-Waldhaus vorhandenen Thierarten.

| Name | Aufenthalt und sonstige Bemerkungen. |
| --- | --- |
| Dachs | Kommt nur selten und zunächst Ludwigsthal und Lindberg vor. |
| Steinmarder | deßgleichen |
| Edelmarder | Kommt im ganzen Revierbezirke ziemlich häufig vor. |
| Fischotter | Kommt zwar auf sämtlichen Triftbächen des Reviers aber selten vor. |
| Reh | Überall und häufig |
| Auerhuhn | Kommt größtentheils in den Auen und nicht gar häufig. |
| Haselhuhn | Im ganzen Revier und ziemlich häufig |
| Schnepfe | deßgleichen |
| Von den Fischen | Häufig die Forelle seltener die Asche. |

- 1 -

[S. 4]

B. Beantwortung der Fragen über die Verbreitung der in Bayern einheimischen Baumarten

| Fragen | Beantwortung |
| --- | --- |
| I Welche Baumarten bilden im Revierbezirke die großern reinen, und welche die großern gemischten Bestände? | Ad I. Im diesseitigen Revierbezirke bildet die Fichte die größern reinen, die Tanne, Fichte und Buche hingegen die großern gemischten Bestände. Auf den höher gelegenen Waldarten kommt auch der Ahorn immer aber nur eingesprengt, eben so auch die Lerche vor. |
| II Welche Baume kommen außer geschlossenen Beständen auf lichten Plätzen, Viehweiden, Gemeindegründen u. s. w. vor oder werden daselbst und in den Umfried[un]g[en] und den Grundstücken gezogen? | Ad II. Da der Revierbezirk sich blos auf den ein Continuum bildenden Hochwaldstock erstrecket, also weder Viehweiden noch Gemeinde-Gründe einschließt, so fällt die Beantwortung weg. - |
| III Kommen seltene Baumarten wie z. B. im Gebirge die Lerche | Ad III. Außer den schon ad I. aufgeführten drey Haupt Baumarten, nämlich der Fichte, Tanne und |

---

1845, München o. J., S. 252.

| | |
|---|---|
| die Zirbelkiefer die Eibe p. vor? | Buche, kommen auch noch eingesprengt der Ahorn die Lerche die Linde und die Eibe, letzter zwey aber nur selten vor. |
| IV Findet sich Quercus pedunculata oder Robur oder vielleicht beyde. | Ad IV. Kommen nicht vor. |
| V. Ist Tilia parvifolia oder grandifolia an den Orten wo gewöhnlich Linden gepflanzt werden, im Gebrauch, und wie gedeihen sie. | Ad V. Da im dießseitigen Revier noch niemals Linden gepflegt wurden so kann über das Gedeihen derselben auch nichts angegeben wohl aber die Vermuthung ausgesprochen werden, daß erstere Gattung dem hiesigen Klima und Boden am besten zutragen dürfte. |

- 2 -

[S. 5]

| Fragen | Beantwortung |
|---|---|
| VI. Welche Stauden sind auf vorkommenden Haiden[61] oder Moorgründen vorherrschend? | Ad VI Haiden oder Moorgründe sind im dießseitigen Reviere nicht vorhanden. |
| VII. Finden sich einzeln durch ihre Dimensionen oder ihr alter besonders ausgezeichnete Individuen von Baumarten | Ad VII Tannen und Fichten zu 6 1/2 bis 7 bay. Fuß Durchmesser[63] bei Brusthöhe und einer Länge von 170 - 190 Fuß[64] sind im hiesigen Reviere keine seltene Erscheinung; Auch finden sich einzelne Tannen, welche weit über 300 Jahre zählen und zum größten Theile noch festes Holz liefern. Eine kleine halbe Stunde oberhalb der Einöde Zwieslerwaldhaus auf ein gegen die Sonne geschützten Standorte steht eine noch gesunde Tanne, welche bei Brusthöhe 7 |

---

[61] Der Begriff ‚Haide' entspricht vermutlich der heutigen Schreibweise ‚Heide' und stellt eine ältere oder regional-bayerische Variante des Wortes dar.

[63] Ein Meter beträgt 3,42631 bayerische Fuß, damit handelt es sich bei einem bayerischen Fuß um 0,29186 Meter. Demnach beläuft sich der Durchmesser dieser Tannen und Fichten auf 1,89709 m bis 2,04302 m.

Für die Umrechnung, siehe: Grebenau, Heinrich: Tabellen zur Umwandlung des bayerischen Masses u. Gewichtes in metrisches Maß und Gewicht und umgekehrt: nebst dazu gehörigen Preisverwandlungen. Auf Grund der mit allerhöchster Verordnung vom 13. August 1869 amtlich bekannt gemachten Verhältnißzahlen. Mit einer kurzen Geschichte und den nöthigen Erläuterungen des metrischen Mass- und Gewichts-Systems. Mit 1 lith. Tafel (München 1870), Tabelle 2.

[64] 49,62 m bis 55,45 m.

| Buchen[62] | 1/2 Fuß[65] mißt und etwa gegen 300 Jahre zählen mag: ein aehnliches Exemplar befindet sich auch am Gipfel des Hochberges. Auch Buchen zu 3 Fuß Durchmesser[66] bei Brusthöhe und einer Länge von 130[67] faßen sind nicht selten, leider sind aber die meisten starken Buchen nicht mehr gesund in der Regel schon anbrüchig. Traditionen u. Volkssagen über derlei ausgezeichnete Baume sind nicht bekannt. |
|---|---|
| | Zwieslerwaldhaus den 29 Augst. 1845 gehorsamster Seninger K. Revfstr. |

- 3 -

[S. 6]

[Bericht des Revierförsters zu Hohenbogen Friedrich v. Krafft-Festenberg, 1845, September 2]

B.
      Fragen über die Verbreitung der in Bayern einheimischen Baumarten.

| I. Welche Baumarten im Amtsbezirke bilden die größern reinen und welche die größern gemischten Bestände? | Ahorn, /: acer pseudoplatanus et platanoides :/ Buchen /: fagus sylvatica :/, dann Tannen Fichten und Birken bilden die größern reinen, und auch, exclus: die Birke, die größern gemischten Bestände. |
|---|---|
| II. Welche Bäume kommen ausser geschlossenen Beständen auf lichten Plätzen, Viehweiden p p vor, oder | Auf freyen lichten Plätzen, namentlich Vieh- weiden kommt häufig die Erle /: betula alnus :/[68] und sehr allgemein die Birke /: betula alba :/[69] vor. |

---

[62] Die Nennung von ‚Buchen' erfolgte hier durch eine nachträgliche Ergänzung mit Bleistift.

[65] 2,19 m.

[66] 0,88 m.

[67] 37,94 m.

[68] Die Bezeichnung *Betula alnus* ist taxonomisch nicht korrekt, da *Alnus* als Gattungsname stets an erster Stelle stehen muss, beispielsweise bei *Alnus incana* (Grauerle). Sowohl Birke (*Betula*) als auch Erle (*Alnus*) gehören jedoch zur Familie der Birkengewächse (*Betulaceae*), was darauf hindeuten könnte, dass die damalige Klassifikation abweichend war. In Deutschland ist *Alnus glutinosa* (Schwarzerle) die häufigste Erlenart, weshalb es möglich ist, dass sich die Angabe darauf bezieht.

[69] Die Bezeichnung *Betula alba* ist taxonomisch nicht nachweisbar. Innerhalb der Gattung *Betula* (Birken) ist in Mitteleuropa insbesondere *Betula pendula* (Hänge-Birke) verbreitet, weshalb sich die Angabe möglicherweise darauf beziehen könnte. Da jedoch lediglich von ‚Birke' gesprochen wird, lässt sich mit Sicherheit nur auf die Gattung schließen.

| | |
|---|---|
| werden daselbst und in Umfriedungen der Grundstüke gezogen? | Die Sommerlinde gewöhnlich bey Kirchen von ansehnlicher Größe und Stärke. In umfriedeten Grundstücken vorzugsweise, jedoch nicht sehr häufig der Kirsch- Zwetschgen- Birn- und Apfelbaum in nicht veredelter Art. |
| III. Kommen seltene Baumarten, wie z. B. im Gebirge die Lerche, Zirbelkiefer, Eibe, im Flachlande Pyrus torminalis, Sorbus domestica p p vor? | Der Eibenbaum erscheint nur selten in Tannen- und Fichten-Waldungen der mässig hohen Berghänge, worunter ein Exemplar von seltner Stärke, namlich 60' lang und 9'' mittl. Durchmesser[70] vorkommt. /: Kötztingerberg am großen Regen nördlich. Hannge[71] :/ |

- 1 -

[S. 7]

| | |
|---|---|
| | Häufiger erscheint die Esche /: fraxinus exelsior :/ von sehr schönen Wuchse und Wachstum gemischt mit Ahorn und Buche in dem südlichen Gehänge des Hohenbogen und Kaitersberges. Auch die Lerche zeigt in den höhren Lagen gutes Fortkommen. |
| IV. Findet sich Quercus pedunculata oder Robur,[72] oder vielleicht beyde und beyläufig in welchem Verhältnisse? | In den Niederungen findet sich quercus robur weniger pedunculata jedoch nur sehr Einzeln auf Wiesen und sogenannten Haus-Angern in einem höchst unbedeutendem Verhältnisse. |
| V. Ist tilia parvifolia oder gardifolia an den Orten, wo gewöhnlich Linden gepflanzt werden im Gebrauch und wie gedeihen sie? | Nur tilia europaea[73] kommt einzeln in Ortschaften und bey Kirchen jedoch von schönem Wuchs vor, welche ihr daseyn den Pflanzungen der Vorzeit zu danken hat, neuere Pflanzungen sind selten und mit der nemlichen Spezies. |

---

[70] Es ist anzunehmen, dass es sich bei ‚60'' um 60 bayerische Fuß und bei ‚9''' um 9 bayerische Fuß handelt. Entsprechend hätte dieser Eibenbaum wohl eine Höhe von 17,52 m erreicht und einen mittleren Durchmesser von 2,63 m, was für diese Baumart eine außergewöhnliche Größe und Breite darstellt.

[71] Hier ist wohl der nördliche Hang des Kötztinger Berges gemeint.

[72] *Quercus robur* bezeichnet die Stieleiche, wobei *pedunculata* offenbar als Synonym für *robur* verwendet wurde. Im vorliegenden Kontext scheinen die beiden Namen jedoch zur Unterscheidung verschiedener Arten genutzt worden zu sein, obwohl unklar bleibt, welche spezifische Bedeutung pedunculata in diesem Fall haben könnte.

[73] *Tilia europaea* bezeichnet eine Hybride aus Sommer- (*Tilia platyphyllos*) und Winterlinde (*Tilia cordata*).

| VI. Welche Sträucher sind auf vorkommenden Haiden- oder Moor-Gründen vorhanden? | Moorgründe bestehen in größerer Ausdehnung nicht, und werden schon zur Grasnutzung benutzt. Auf Haide-Gründen sind vorherrschend: 1. der gemeine Hartriegel /: ligust: valgare :/[74] 2. Faulbeerbaum /: rhamnus frangula :/[75] 3. Schlingenbeutel /: virburum apulus :/[76] |
|---|---|

- 2 -

[S. 8]

| | 4. Besenpfrieme /: spartium scoparium :/[77] 5. gemeine Wachholder /: juniperus communis :/ 6. Brombeerstrauch /: rubus fruticosus :/[78] 7. gemeine Heidelbeere /: vacc: myrtillus :/[79] 8. Preißelbeere /: vacc: vitis idaea :/ 9. gemeine Heide /: erica vulgaris :/[80] 10. auch die Alpenrose /: rosa alpina :/[81] . - |
|---|---|
| VII. Finden sich einzeln durch ihre Dimensionen auf ihr Alter besonders ausgezeichnete Individuen von Baumarten? | Buchen, Tannen und Fichten zu 90' lang[82] und 3 bis 5' Stock-Durchmesser[83] finden sich nicht selten als Hauptbesta[n]ds- Masse zu circa 200 bis 300 Jahre alt vor. Von Tilia europaea sind zwey ausgezeichnete Exemplare in den Ortschaften Ried und Gotzendorf an nordwestlichen nicht geschüzten Hängen, wovon die Stärkere 16 Fuß im Stockdurchmesser[84] hat und circa 300 Jahre alt ist, im Innern durchaus bis zu einer Höhe von 15' hohl[85] und wird zu einer |

---

[74] Der Begriff ‚Gemeiner Hartriegel' ist taxonomisch nicht eindeutig nachweisbar. Vermutlich ist damit der Rote Hartriegel (*Cornus sanguinea*) gemeint, eine in Deutschland häufig vorkommende Art. *Ligustrum vulgare* (Gewöhnlicher Liguster) wird ebenfalls genannt, gehört jedoch nicht zur Gattung *Cornus* und steht daher taxonomisch in keinem Zusammenhang mit dem Hartriegel.

[75] Der Begriff bezieht sich heute auf den Faulbaum (*Frangula alnus*), wobei der wissenschaftliche Name unverändert geblieben ist.

[76] Vermutlich ist hier der Gewöhnliche Schneeball (*Viburnum opulus*) gemeint.

[77] Dieser Name ist eine ältere Bezeichnung für den Besenginster, der heute als *Cytisus scoparius* bekannt ist.

[78] Dieser Name bezeichnet einen Artkomplex, der zahlreiche Arten umfasst, darunter auch Brombeerarten.

[79] Hierbei handelt es sich um Vaccinium myrtillus.

[80] *Erica* bezeichnet die Gattung der Heidekräuter, allerdings konnte vulgaris nicht eindeutig zugeordnet werden. Eine mögliche heutige Bezeichnung könnte *Erica carnea* (Schneeheide) sein, die in Süddeutschland häufig vorkommt. Es ist jedoch denkbar, dass der Name *vulgaris* historisch verwendet wurde.

[81] Die Alpenrose ist vermutlich "*Rosa pendulina*" (heute: Gebirgsrose) *Rosa alpina* ist wohl aber ein Synonym, das früher verwendet wurde.

[82] 26,27 m.

[83] 0,88 m bis 1,46 m.

[84] 4,67 m.

| | Kapelle benuzt; die Krone vielästig und gesund. Kötzting am 2ten September 1845 v Krafft Revierförster |
|---|---|

- 3 -

[S. 9]

A. Verzeichnis der Thierarten, von deren Vorhandenseyn und Wohnort Nachricht gewünscht wird.

| Namen | Wohnort und sonstige Bemerkungen. |
|---|---|
| Steinmarder | Kommt allenthalben in den Ortschaften des Revierbezirks vor. |
| Edelmarder | Häufig in den Hochwaldungen der Ossa, Hohenbogen u. Keitersberg. |
| Fischotter | Nicht selten im schwarzen und weisen Regenflusse und deren Nebenbächen. |
| Edelhirsch | Nur durchwandernd aus dem benachbarten Böhmen, jedoch nicht selten. |
| Reh | In den Hochwaldungen des Hohenbogens und Haidsteines häufig. |
| Saatkrähe | Nur in den niedern Lagen, jedoch nicht sehr häufig. |
| Nachtigall | In den Niederungen, jedoch nur selten. |
| Auerhahn | In den Hochwaldungen des Arber, Ossa, Hohenbogen u. Keitersberg nicht selten. |
| Birkhuhn | Nicht häufig in den Vorbergen der höhern Gebirge. |
| Haselhuhn | In den höhern Gebirgen des Arber, Ossa, Hohenbogen, Keitersberg und Haidstein sehr häufig. - |
| Schnepfe | Brütet sehr häufig in den Hochwaldungen dieser Gebirge. |
| Von Fischen | Im schwarzen Regenflusse Huchen, Schiele, Aschen und Hechte. Im weißen Regenfluße Forellen, welche auch durchaus sehr häufig in den Gebirgsbächen des Arber, Ossa und Keitersberg vorkommen. In kleinen Nebenbächen finden sich auch Aale. |
| | Kötzting am 2ten September 1845 v Krafft Revierförster |

- 4 -

[Bericht des Revierförsters zu Bodenmais Karl Frhr. v. Asch, 1845, September 5]

[S. 10]

    Verzeichnis der im Bezirke des k. Bergamts Forstrevier Bodenmais vorkommenden Thierarten.

---

[85] 4,38 m.

| Name. | Wohnort und sonstige Bemerkungen. | Name. | Wohnort und sonstige Bemerkungen. |
|---|---|---|---|
| Säugethiere.[86] | | Vögel. | |
| Dachs | sehr selten | Uhu | wenige |
| Steinmarder | mittelmäßig | Alpendohle | mittelmäßig. |
| Edelmarder | do | Auerhuhn | wenig. |
| Fischotter | einige im Schwarzbach. | Birkhuhn | sehr selten |
| Luchs | wenige | Haselhuhn | mittelmäßig |
| Rehe | wenige | Waldschnepfe | do |
| Haasen[87] | mittelmäßig. | Bekassine | strichweise. |
| Amphibien. | | Fische. | |
| Kupferotter[88] | Sehr wenige | Forelle | in den Bergbächen, nicht zu häufig. |
| | | Asche | im Regenfluß, nicht zu häufig |
| | | Huche | im Regenfluß nicht zu häufig, letzterer von ziemlicher Größe. - |
| | | Bodenmais am 5ten September 1845. Frhr. v. Asch K. RFster | |

- 5 -

[S. 11]

## Verbreitung der in Bayern einheimischen Baumarten

| I. Welche Baumarten im Amtsbezirke bilden die größeren reinen, und welche die größern gemischten Bestände? | In den Staatswaldungen kommen größere reine Bestände nicht vor; die gemischten größeren Bestände bilden Fichten, Tannen und Buchen; in den Privatwaldungen bilden die Birken die |
|---|---|

---

[86] Die Zwischenüberschriften („Säugethiere", „Vögel", „Amphibien" und „Vögel") sind im ursprünglichen Fragebogen nicht vorgegeben und wurden eigenständig durch den Revierförster ergänzt.

[87] Der Hase wird im Fragebogen nicht explizit abgefragt; seine Nennung erfolgt als selbstständige Ergänzung des Revierförsters.

[88] Die Kupferotter, heute als Kreuzotter (*Vipera berus*) bekannt, gehört systematisch zu den Reptilien, wurde jedoch irrtümlich unter der Kategorie Amphibien aufgeführt.

| | größeren reineren Bestände. |
|---|---|
| II. Welche Bäume kommen ausser geschlossenen Beständen auf lichten Plätzen Viehweiden, Gemeinde-Gründen u. s. w. vor, oder werden daselbst und in den Einfriedungen der Grundstücke gezogen? | Buchen, Birken, Pappeln, Linden, Ahorn, Elzbeerbäume, Föhren, Fichten[,] Tannen, Lerchen, Vogelbeerbäume, Eschen, Eichen, Schwarz- und Weißerlen. |
| III. Kommen seltene Baumarten wie z. B. im Gebirge die Lerche, Zirbelkiefer, Eibe, im Flachlande Pyrus torminalis, Sorbus domestica u. a. vor? | Kommt nur die Lerche vor, und zwar sehr wenig, von Eiben findet sich keine Spur mehr. |

- 6 -

[S. 12]

| IV. Findet sich Quercus pedunculata oder Robur, oder vielleicht beyde und beyläufig in welchem Verhältnisse? | Kommen nur vereinzelt und sehr selten auf Privat- und Gemeindegründe vor. |
|---|---|
| V. Ist tilia parvifolia oder grandifolia an den Orten, wo gewöhnlich Linden gepflanzt werden im Gebrauche, und wie gedeihen sie? | Sind sehr wenig im Gebrauche; doch gedeihen jene welche vorkommen, sehr gut. - |
| VI Welche Sträucher sind auf vorkommenden Haiden oder Moorgründen vorherrschend? | Waiden, Wachholder, Haselnuß, Schlehdorn, Him- Brom- Heidel- Preusel- beeren, Faulbeerstrauch. |
| VII. Finden sich einzelne durch ihre Dimensionen und ihr Alter besonders ausgezeichnete Individuen von Baumarten?<br>Man wünscht hierbey ausser der Bestimmung der Spezies möglich genaue Angaben des Standortes, der Dimensionen, des muthmaßlichen Alters und der allenfalls über solche Individuen gängigen Traditionen u. Sagen? | Kommt nichts ausgezeichnetes vor. |
| | Bodenmais am 5ten Sept[em]ber 1845.<br>Fr[ei]h[er]r v Asch k. R[evier]F[ör]ster |

- 7 -

[Bericht des Revierförsters zu Zwiesel Max Ney, ohne Datum]

[S. 13]

|  | Ad B. |
|---|---|
| I, Welche Baumarten im Amtsbezirke bilden die größeren reinen, u. welche die größeren gemischten Bestände? - | Die Fichte bildet die größeren reinen, und die Tanne und Buche mit der Fichte bilden die größeren gemischten Bestände, in den sog. Birkenbergen erscheint auch die Birke ganz rein. - |
| II, Welche Bäume kommen außer geschloßenen Beständen auf lichten Plätzen, Viehweiden pp. vor, oder werden daselbst u. in den Umfriedungen der Grundstücke gezogen? | Außer geschloßenen Beständen auf lichten Plätzen p.p. erscheint hier vorzugsweise die Birke, auch die Fichte u. hie u. da die Föhre. - |
| III, Kommen seltenere Baumarten, wie z. B. im Gebirge die Lerche, die Zirbelkiefer, im Flachlande pyrus torminalis, sorbus domestica p. vor? | Zu den selteneren Baumarten gehören hier die Eibe, Esche u. Ahorn. - |
| IV. Findet sich quercus pedunculata oder robur, oder vielleicht beide, u. in welchem Verhältnisse? | Kommt nicht vor. - |
| V. Ist tilia parvifolia oder grandifolia an den Orten, wo gewöhnlich Linden gepflanzt werden im Gebrauche, u. wie gedeihen sie? - | In hiesiger Gegend kommt tilia grandifolia vor; sie wurde, wie die vorhandenen, einzelnen Exemplare zeigen, fast in allen Ortschaften gepflanzt u. steht im üppigen Wuchse; im Hochwalde jedoch erscheint sie selten u. in nicht besonderen Gedeihen. - |

- 10 -

[S. 14]

| VI, Welche Sträucher sind auf vorkommenden Haiden oder Moorgründen vorherrschend? [89] | Hier erscheinen vorzüglich der Wachholderstrauch, der Haselstrauch[90] d. mehrere Weidenarten, z. B. die Salweide p. - |
|---|---|
| VII, Finden sich einzeln durch ihre Dimensionen u. ihr Alter besonders ausgezeichnete Individuen von Baumarten p. ? - | Kommt nichts vor. - |

---

[89] „Sehr mangelhaft!" erfolgte hier durch eine nachträgliche Ergänzung mit Bleistift. und bezieht sich wohl auf die Moorgründe.

[90] „oft!"[?] wurde nachträglich mit Bleistift hinzugefügt und bezieht sich wohl auf den Haselstrauch.

[S. 15]

### Revier Zwiesel
### ad. A.

| Namen | Wohnort u. sonstige Bemerkungen. |
|---|---|
| Bär | - |
| Dachs | Kommt selten vor. |
| Steinmarder | - |
| Edelmarder | Kommt in den hiesigen Waldungen sehr häufig vor. |
| Fischotter | Ist in den hiesigen Bächen keine Seltenheit. - |
| Luchs | - |
| Wildkatze | - |
| Wolf | - |
| Biber | - |
| Hamster | - |
| Murmelthier | - |
| Wildschwein | - |
| Edelhirsch | Nur als Wechselwild; übrigens auch selten. |
| Dammhirsch | - |
| Reh | Ist in den hiesigen Waldungen zu Hause; kommt also häufig vor. |
| Gemse | - |
| Lämmergeier | - |
| Steinadler | - |
| Seeadler | - |
| Fischaar | - |
| Uhu | - |
| Saatkrähe | Kommt häufig vor, zieht aber im Winter fort. |
| Alpendohle | - |
| Steinkrähe | - |
| Nachtigall | - |
| Mauerspecht | - |
| Auerhuhn | Kommt häufig vor, und brütet in den hiesigen Waldungen. |
| Birkhuhn | Ist jetzt sehr selten; obschon vor einem Jahrzehend häufig vorkommend. |
| Haselhuhn | Ist hier zu Hause u. brütet in diesen Waldungen. |
| Schneehuhn | - |

- 8 -

[S. 16]

| Namen | Wohnort u. sonstige Bemerkungen |
|---|---|
| Fasan | - |
| Trappe | - |
| Schnepfe | Brütet in den hiesigen Waldungen, u. kommt demnach sehr häufig vor. |
| Beccaßine | - |
| Kranich | - |
| Rohrdommel | - |
| Weißer Storch | - |
| Schwarzer Storch | - |
| Höckerschwan | - |
| Singschwan | - |
| Gemeine Wildgans | Sehr selten, u. nur als Strichvogel. |
| Saatgans | - |
| Enten | Sehr selten u. nur als Strichvogel. |
| Kupferotter | - |
| Von Fischen die merkwürdigsten. | Forelle, Asche, Huche, Hechten. |

- 9 -

[S. 17]

Ad Num. 1783.　　　　　　　　　[...] 170.　　　Reichenach den 11ten September 1845.
E[xpeditio] No. 1873　　　　　　　　　　　Praes[entiert]. am 15 Septb 1845
Bericht des königl. Forsteyförsters von Weißenstein
Königliches Forstamt Zwiesel!
Die geographische Verbreitung der Thier- und Pflanzenarten in Bayern be[treffend].
In Erledigung des verehrlichsten Auftrages vom 28ten praes. 31ten vorigen Monats, werden in Nachstehenden die gemäß hoher Regierungs-Entschließung über rubr. Betr. gegegebenen Fragen mit möglichster Genauigkeit beantwortet.
A.
die im Forsteybezirke Weißenstein vorhandenen Thierarten
a. Säugethiere
1. Steinmarder und 2. Edelmarder, kommen allenthalben im Forsteybezirke vor, so auch 3. der Fischotter, dann 4. Rehe nur wechselweise, aber sämtlich vorstehende Thierarten nicht häufig.
b. Vögel
5. das Auerhuhn und 6. das Haselhuhn kommen in den größern Waldungen des Forsteybezirkes vor, erste in geringerer Anzahl /: im Privatwald Hollerberg bey Eberhartsried, im Privatholz Baiken unweit Weißenstein und den Rinchnamünderhölzer:/

letztere häufiger /: im kgl. Walddistrikt
<div align="center">- 12 -</div>

[S. 18]

Hollerberg und Weissensteinerau :/ 7. Die <u>Schnepfe</u> kommt allenthalben vor, wo sie auch wie die vorigen Vogelarten brütet, jedoch nicht häufig. 8. Die <u>Bekaßine</u>, strichweise im Frühjahr und Herbste auf ausge-dehnten Bachwiesen /: namentlich bey Reichnach :/ jedoch in geringer Anzahl. 9. Von den <u>Enten</u> brütet fast alle Jahre ein oder zwey Paare Stockenten an der Rinchnacher Ohe, und es kommen im Frühjahr und Herbste strichweise mehrere Entenarten aber nicht häufig vor.
<u>c. Fische</u>
10. Die <u>Forelle</u> kommt nur in den kleinen Bächen und in den grössern, soweit dieselben steinigt und starkes Gefäll haben, vor /: Rinchnach Bach, Kühbach, Siegbächl, Haidbächel, Schloßauer Ohe :/ 11. <u>Aschen</u>, allenthalben in den größern Bächen /: Regenfluß, Rinchnacher- und Schloßauer Ohe :/ 12. <u>Hechte</u> ebenfalls in vorigen Gewässern. 13. <u>Huchen</u> aber nur im Regenflußße und Schloßauerohr.
<div align="center">- 13 -</div>

[S. 19]

<u>B.</u>
<u>die im Forsteybezirke Weissenstein einheimischen Baumarten.</u>
<u>Ad Frage I.</u>
Die Fichte bildet die größern reinen, dann Fichte und Tannen die größern gemischten Bestände.
<u>Ad Frage II.</u>
Außer geschloßenen Beständen auf lichten Plätzen, Viehweiden und Gemeindegründen, kommen vor die Birke, Erle, die Fichte aber größtentheils in schlechten Zustande, und die Föhre - überall der Wachholderstrauch. In den Umfriedigungen der Grundstücke, kommen einzeln die Esche /: Fraxinus excelsior :/ die Rüster, der Ahorn, die Zitterpappel dann die Linde, und in schwachen Exemplaren die Eiche vor.
<u>Ad. Frage III.</u>
Als seltene Baumart kommt die Lärche einzeln an Hausgärten in mitunter schönen Exemplaren vor.
<u>Ad Frage IV.</u>
Quercus pedunculata und robur, finden sich sehr selten und mit geringen Dimensionen auf den Feldrainen und in Gehöften.
<u>Ad Frage V.</u>
Tilia parvifolia und grandifolia finden sich auf Dorfängern und bey Einzelhöfen öfters und gedeihen sehr gut.
<div align="center">- 14 -</div>

[S. 20]

<u>Ad Frage VI.</u>

Auf Haiden ist der Wachholder vorherrschend; Moorgründe werden als Wiesen benützt.
Ad Frage VII.
An der Einöde Ebenhof 1/4 Stunde südwestlich von Weißenstein findet sich eine bemerkenswerthe Linde, titlia grandifolia. Diese misst an Brusthöhe 22 Fuß im Umfange[91], theilt sich bey einer Schafthöhe von 10 Fuß[92] fast regelmäßig in 10 Aeste die bis zu 6' Umfang[93] haben und einen Kreis von 235 Fuß Peripherie[94] beschatten, während das majestätisch aufsteigende Gipfelstück sich wieder in 7 Aeste theilet die einen Durchmesser von 1 1/2 bis 2 Fuß[95] haben, und der Baum eine Höhe von 70 bis 80 Fuß[96] erreicht. Ueber das Alter läßt sich nichts zuverlässiges angeben, mag aber immer mehr denn 300 Jahre zählen. Der Stamm ist frisch und gesund und hängt voller Saamen.
Einem Königlichen Forstamte
gehorsamster
Pfisterer
Kgl. Forsteyförster

- 15 -

[S. 21][97]

Ad Num 16671    Exp[editio] No 1783        Zwiesel den 17 Sept 1845
An die Königl. Regierung von Niederbayern K[ammer] d[er] F[inanzen] in Landshut
die geographische Verbreitung der Thier- und Pflanenarten in Bayern betr[effend].
Schmid[t?][98]

Im Nachgange hoher Reg[ie]r[un]gs Entschließung vom 18/23 v[origen]. M[ona]ts. No 16671 rubr betr[effend] werden die anher gegebenen 2 Schemata A und B, nach vorgenomener Ausfüllung der Rubriken, wieder in unterthänigster Vorlage gebracht
Ehrfurchtsvollst empiehlt
sich
Euer unterthänigst gehorsamstes Forstamt
Klein

---

[91] 6,42 m.

[92] 2,92 m.

[93] 1,75 m.

[94] 68,59 m.

[95] 0,44 m bis 0,58 m.

[96] 20,43 m bis 23,35 m.

[97] Seite 21 wurde nicht mit Bleistift nummeriert.

[98] Vgl. das Hof= und Staats=Handbuch des Königreichs Bayern 1845, München o.J., S. 252, wo unter den „Räthe[n]" der „Kammer der Finanzen" der Regierung des Kreises Niederbayern in Landshut „Franz Schmid, Forstinspector, K[öniglicher] Forstrath, Ehrenkreuz des Ludwig=Ordens" geführt wird.

[S. 22]

A.
ad Num 12613
Verzeichnis der Thierarten, von deren Vorhandensein u. Wohnart, Nachricht gewünscht wird.
Bemerkung: Beschränkt sich die Verbreitung nur auf gewiße Ortlichkeiten, so ist der nächste Ort oder Ortschaft in Parenthese /: :/ einzuschalten. Bey den Vögeln sind nur solche aufzuführen, welche im Bezirke selbst brüten oder doch, den Winteraufenthalt daselbst nehmen, oder die zur Zeit Gegenstand der Jagd sind. – Anzugeben ist auch, ob die Thierart häufig oder selten ist. -

| Namen | Wohnort u. sonstige Bemerkungen |
|---|---|
| Bär | - |
| Dachs | Kömmt sehr selten vor. [Angabe in Parenthese durchgestrichen, nicht leserlich] |
| Steinmarder | Kömmt in vielen Ortschaften jedoch nicht sehr häufig vor |
| Edelmarder | Kömmt in den Waldungen sehr häufig vor |
| Fischotter | Findet sich auf den Flüssen u. Bächen jedoch nicht sehr häufig |
| Luchs | - |
| Wildkatze | - |
| Wolf | - |
| Biber | - |
| Hamster | - |
| Murmelthier | - |
| Wildschwein | - |
| Edelhirsch | Dann und wann als Wechselwild |
| Dammhirsch | - |
| Rehe | In den größeren zusammenhängenden Waldungen sehr häufig |
| Gemse | - |

- 16 -

[S. 23]

| Namen | Wohnort u. sonstige Bemerkungen |
|---|---|
| Lämmergeyer | - |
| Steinadler | - |
| Seeadler | - |
| Fischaar | - |
| Uhu | Höchst selten vorkommend |
| Saatkrähe | - |
| Alpendohle | - |
| Steinkrähe /: | - |
| corvus graculus :/ | |
| Nachtigall | - |
| Mauerspecht /: Tichodroma muraria :/ | - |
| Auerhuhn | In den größeren Waldungen ziemlich häufig vorkommend |
| Birkhuhn | Sehr selten vorkommend |
| Haselhuhn | In den größern Waldungen ziemlich häufig vorkommend |
| Schneehuhn | - |
| Fasan | - |
| Trappe | - |
| Schnepfe | Brütet in den hiesigen Waldungen und kommt ziemlich häufig vor |
| Bekassine | Kommt selten und nur strichweise vor |
| Kranich | - |
| Rohrdommel | - |
| Weißer Storch | - |
| Schwarzer Storch | - |

- 17 -

[S. 24]

| Namen | Wohnort u. sonstige Bemerkungen |
|---|---|
| Höckerschwan | - |
| Singschwan | - |
| Gemeine Wildgans | - |
| Saatgans | - |
| Enten | Sehr selten vorkommend. |
| Kupferotter | -[99] |
| Von Fischen die merkwürdigsten | Die Steinforelle, [Angabe durchgestrichen, nicht leserlich] die Aesche /: Asche :/ der Hecht sind die in den hiesigen Gewässern am häufigsten vorkommenden Fische. Huchen /: Hauchforelle :/ und Schiele[100] kommen seltner vor. [Angabe durchgestrichen, nicht leserlich] |
| | Zwiesel den 17 Sept 1845<br>Königl Bayr Forstamt Zwiesel<br>Klein |

- 18 -

---

[99] „1902 Verhältnismäßig häufig; damals noch nicht? darum hat sich eben niemand gekümmert!-" erfolgte hier durch eine nachträgliche Ergänzung mit Bleistift.

[S. 25]

<u>B.</u> <u>I.</u>
<u>Ad Num: 12613.</u>
<u>Fragen</u> über die Verbreitung der in Bayern einheimischen Baumarten.
<u>Bemerkung:</u> Der Standort seltener Baumarten oder besonders merkwürdigen Spezies ist durch Beifügung der nächsten Ortschaft in Parenthese /: :/ zu bezeichnen.

I. Welche Baumarten im Amtsbezirk bilden die größern reinen und welche die größern gemischten Bestände?
Die Fichte, Föhre, Rothbuche und Birke bilden hie und da größere reine Bestände. Die größern
gemischten Bestände werden aber von der Fichte Tane und Buche gebildet, davon hie und da auch der Ahorn /: Drey- oder Spitzahorn :/ die Esche und die Ulme als edlere, die Birke, die Aspe und die Erle /: Schwarz- u. Weißerle :/ so wie die Fohre als unedlere Holzarten beygemischt sind.
II. Welche Bäume kommen außer geschlossenen Beständen auf lichten Plätzen, Viehweiden, Gemeinde-Gründen u. s. w. vor, oder werden daselbst und in den Umfriedungen der Grund-
stücke gezogen?
Die Wintereiche (quercus robur) die Esche, die Schwarz- und Weißerle, die Zitterpappel, die Birke, die Linde, die Buche, der Ahorn, der Vogelbeerbaum, die Föhre die Fichte die Tane, die Lärche.

- 19 -

[S. 26]

<u>II.</u>
III. Kommen seltenere Baumarten, wie z. B. im Gebirge, die Lerche, die Zirbelkiefer, Eibe; im Flachlande Pyrustorminalis, Sorbus domestica u. a. vor?
Die Lärche wird in den Staatswaldungen durch Saat und Pflanzung herangezogen, die Eibe findet sich nur mehr sehr selten vor.
IV. Findet sich Quercus pedunculata oder Robur, oder vielleicht beide und beiläufig in welchem Verhältnisse?
Die Eiche komt nur sparsam auf Privat und Gemeindegründen vor, und zwar meistens die Quercus robur
V. Ist tilia paroifolia oder grandifolia an den Orten, wo gewöhnlich Linden gepflegt werden, im Gebrauche, und wie gedeihen sie?
In hiesiger Gegend kömt die Tilia grandifolia als gepflanzter Baum in allen Ortschaften vor und gedeiht ganz vorzüglich gut.

- 20 -

[S. 27]

---

[100] Hier ist wohl die „Schleie" gemeint.

III.
VI. Welche Sträucher sind auf vorkommenden Haiden oder Moorgründen vorherrschend?
Auf Haiden ist der Wachholderstrauch und auf Moorgründen sind die Vaccinien vorherrschend.
VII. Finden sich einzeln durch ihre Dimensionen und ihr Alter besonders ausgezeichnete Individuen von Baumarten? Man wünscht hiebei außer der Bestimmung der Spezies möglichst genaue Angaben des Standortes, der Dimensionen, des muthmaßlichen Alters und der allenfalls über solche Individuen gängigen Traditionen und Sagen?
In der Ortschaft Ried Landgerichts Kötzting kommt eine Linde vor /: Tilia grandifolia :/ deren Stockdurchmesser 16 Fuß[101] beträgt. Dieser Baum ist bis zu einer Höhe von 15 Fuß[102] hohl und wird zu einer Kapelle benützt. Die Krone ist vielästig und gesund. Das Alter mag wohl 300 Jahre betragen. Zu Einöd Ebenhof 1/4 Stunde südwestlich von Weissenstein Landgerichts Regen befindet sich eine Tilia grandifolia von 7 Fuß[103] Stockdurchmesser. Dieser Baum theilt sich bey 10 Fuß[104] Schafthöhe in 10 Aeste, die bis zu 2 Fuß[105] durchmesser

- 21 -

[S. 28]

haben, und überschirmt eine Fläche von 75 Fuß[106] Durchmesser. Das mehrastlich aufsteigende Gipfelstück theilt sich wieder in 7 Aeste von 1 1/2 bis 2 Fuß[107] Stärke und die ganze Höhe des Baumes beträgt gegen 80 Fuß[108]. Das Alter möchte zu 300 Jahre angenommen werden dürfen. Der Baum ist frisch und gesund, und trägt reichlich Saamen.
Tanen und Fichten u 6 bis 7 bayr. Fuß[109] Durchmesser bey Brusthöhe und einer Höhe von 160 bis 190 Fuß[110] sind in den Waldungen des Revier Zwieslerwaldhaus und Zwiesel keine seltene Erscheinung. Das Alter solcher Stäme beträgt 3 bis 4 hundert Jahre.
Zwiesel den 17 Sept 1845.
Königl Bayr Forstamt Zwiesel
Klein

---

[101] 4,67 m.

[102] 4,38 m.

[103] 2,04 m.

[104] 2,92 m.

[105] 0,58 m.

[106] 21,89 m.

[107] 0,44 m. bis 0,58 m.

[108] 23,35 m.

[109] 1,75 m. bis 2,04 m.

[110] 46,70 m. bis 55,45 m.